\documentclass[10pt,preprint,epsfig]{emulateapj}

\shorttitle{Local Group distances and publication bias. III. The Small
  Magellanic Cloud} 
\shortauthors{Richard de Grijs and Giuseppe Bono}

\begin{document}

\title{Clustering of Local Group distances: publication bias or
  correlated measurements? III. The Small Magellanic Cloud}

\author{
Richard de Grijs\altaffilmark{1,2} and
Giuseppe Bono\altaffilmark{3,4}
}

\altaffiltext{1} {Kavli Institute for Astronomy and Astrophysics,
  Peking University, Yi He Yuan Lu 5, Hai Dian District, Beijing
  100871, China}
\altaffiltext{2} {Department of Astronomy, Peking University, Yi He
  Yuan Lu 5, Hai Dian District, Beijing 100871, China}
\altaffiltext{3} {Dipartimento di Fisica, Universit\`a di Roma Tor
  Vergata, via Della Ricerca Scientifica 1, 00133, Roma, Italy}
\altaffiltext{4} {INAF, Rome Astronomical Observatory, via Frascati
  33, 00040, Monte Porzio Catone, Italy}

\begin{abstract}
Aiming at providing a firm mean distance estimate to the Small
Magellanic Cloud (SMC), and thus to place it within the internally
consistent Local Group distance framework we recently established, we
compiled the current-largest database of published distance estimates
to the galaxy. Based on careful statistical analysis, we derive mean
distance estimates to the SMC using eclipsing binary systems, variable
stars, stellar population tracers, and star cluster properties. Their
weighted mean leads to a final recommendation for the mean SMC
distance of $(m-M)_0^{\rm SMC} = 18.96 \pm 0.02$ mag, where the
uncertainty represents the formal error. Systematic effects related to
lingering uncertainties in extinction corrections, our physical
understanding of the stellar tracers used, and the SMC's complex
geometry---including its significant line-of-sight depth, its
irregular appearance which renders definition of the galaxy's center
uncertain, as well as its high inclination and possibly warped
disk---may contribute additional uncertainties possibly exceeding
0.15--0.20 mag.
\end{abstract}

\keywords{astronomical databases --- distance scale --- galaxies:
  distances and redshifts --- galaxies: individual (Small Magellanic
  Cloud)}

\section{A Robust Distance to the Small Magellanic Cloud}
\label{intro.sec}

The nearest galaxies in the Local Group contain numerous objects that
can be used to determine robust distances to their hosts. In de Grijs
et al. (2014; Paper I) and de Grijs \& Bono (2014; Paper II), we aimed
at establishing a robust, internally consistent local distance
framework supported by a number of the largest Local Group galaxies
that contain numerous individual distance tracers, including the Large
Magellanic Cloud (LMC), M31, M32, and M33, as well as a number of
well-known dwarf galaxies. Although our statistical treatment of the
individual distance measures to each of these galaxies encountered
unexpected difficulties at some level or another, assigning mean
distances to each galaxy was fairly straightforward. This was
facilitated by either the regular (symmetrical) geometry of the sample
galaxies, their low line-of-sight inclinations, and/or their small
angular sizes.

To date, no such analysis has been performed for the Small Magellanic
Cloud (SMC). Despite its proximity, individual distance estimates to
the galaxy cover a much larger range than those of its larger
neighbor, the LMC. The latter galaxy is often considered a key rung of
the extragalactic distance ladder, and as such robust determination of
its distance has attracted significantly more effort (cf. Paper I)
than the equivalent task pertaining to the SMC. However, this is not
the only reason for the larger scatter in published SMC distance
moduli and its consequently more poorly-known distance.

Schaefer (2008) suggested that the tighter clustering of LMC compared
with SMC distance moduli may be related to sociological effects
(`publication bias') in the distance determination to the LMC
following the publication of the final results of the {\sl Hubble
  Space Telescope} Key Project (HSTKP) on the Extragalactic Distance
Scale (Freedman et al. 2001). He argued that since the SMC was not
included in the HSTKP sample, its ensemble of distance measurements
might be less affected by publication bias. However, in Paper I we
showed that publication bias is unlikely to blame for the tight
clustering of LMC distance moduli over the past two decades. Instead,
we pointed out that improvements in both the quality of the available
data sets---combined with increasing numbers of target objects during
the period of interest---and the theoretical background at the basis
of many methods of distance determination were a more likely
explanation of the convergence in LMC distance moduli.

We believe that this comparison of the Magellanic Clouds is too
simplistic. Schaefer (2008) glossed over a number of important aspects
of the SMC's geometry that make obtaining a clear-cut mean distance
much more challenging for this galaxy than for the LMC. In essence,
the difficulties relate to three aspects. First, the SMC is an
irregular galaxy, exhibiting a bar-like main body with hints of spiral
arms and a very extended `Wing' to the East (for a clear illustration
of the latter, see e.g., Fig. 15 in Sewi{\l}o et al. 2013; see also
Rubele et al. 2015 or http://www.esa.int/spaceinimages/Images/2015/02/
Explor\-ing\_the\_colours\_of\_the\_Small\_Magellanic\_Cloud). This
renders the definition of the galaxy's center troublesome. Few authors
comment on this specifically, although Kochanek (1997), for instance,
states that ``our distances and the Westerlund (1990) value for the
SMC are larger than the Caldwell \& Laney (1991) values because of
differences in defining the Cloud centers.'' More recently, Rubele et
al. (2015) embarked on an exploration of the SMC's spatially resolved
star-formation history, while simultaneously deriving distances to
different areas across the galaxy. They report distances projected
onto both the SMC's kinematic and stellar density centers,
$(m-M)_0^{\rm kin} = 18.97 \pm 0.01$ mag and $(m-M)_0^{\rm stars} =
18.91 \pm 0.02$ mag, which thus implies that one's choice of SMC
center could introduce systematic uncertainties of order 0.05--0.1 mag
in the resulting distance modulus. In this paper, we aim at
determining the `mean' SMC distance to the bulk of its stellar
population, i.e., to a position in the midst of the galaxy's main
body. However, as we will see, the centroids of the different distance
indicators we use vary slightly across the face of the SMC.

Second, the SMC is known to be significantly extended along the line
of sight. Depending on one's tracer and sample selection, the SMC's
depth could be anything from 6--12 kpc (Crowl et al. 2001) up to 20
kpc (Groenewegen 2000; for recent discussions, see e.g., Kapakos \&
Hatzidimitriou 2012; Subramanian \& Subramaniam 2012; Cignoni et
al. 2013; Kalirai et al. 2013; Nidever et al. 2013), although the
Cepheid population associated with the main body implies a shallower
depth of $1.76 \pm 0.6$ kpc (Subramanian \& Subramaniam
2015). Clearly, any tracer population spanning even a fraction of
these reported line-of-sight distances will exhibit a significant
spread in distances which, in turn, will translate into larger
uncertainties and scatter. In addition, selection biases or
small-number statistics will exacerbate the resulting scatter.

Third, whereas the LMC is viewed close to face-on, the SMC's
inclination is much less well-defined and appears to depend on the
stellar tracer (and thus the age of the stellar population) used for
its determination. Based on their analysis of both red clump (RC)
stars and RR Lyrae variables, Subramanian \& Subramaniam (2012)
concluded that the SMC's orientation is almost face-on, characterized
by inclination angles of $i = 0.58^\circ$ and $i = 0.50^\circ$ for the
RC stars (1280 regions, each containing 100 to 3000 RC stars) and RR
Lyrae variables (1904 objects), respectively. Similarly, Haschke et
al. (2012) found a low inclination of $i = 7^\circ \pm 15^\circ$ based
on their sample of 1494 RR Lyrae stars. On the other hand, the large
population of Cepheid variables in the SMC traces a much more highly
inclined disk structure, with inclination estimates ranging from $i =
45^\circ \pm 7^\circ$ (Laney \& Stobie 1986; 23 Cepheids) to $i =
68^\circ \pm 2^\circ$ (Groenewegen 2000; 236 Cepheids), $i = 70^\circ
\pm 3^\circ$ (Caldwell \& Coulson 1986; 63 Cepheids), and most
recently $i = 74^\circ \pm 9^\circ$ (Haschke et al. 2012; 2522
Cepheids). Meanwhile, Rubele et al. (2015) very recently embarked on
near-infrared (IR) color--magnitude diagram (CMD) analysis to derive
$i = 39.3^\circ \pm 5.5^\circ$ for the inclination of the SMC's disk,
with its northeastern quadrant closest to us. They also find that a
warped outer disk (by up to 3 kpc) fits their data best. Careful
geometric corrections of individual objects back to the galaxy's
center will reduce the scatter in the calibration relations, but this
is not always possible.

For instance, let us take $i = 70^\circ$ as an extreme example,
combined with the SMC's size given by de Vaucouleurs et al. (1991), $a
\times b = 9487 \times 5588$ arcsec$^2$ and a `best' distance modulus
to the galaxy of $(m-M)_0 = 18.96$ mag (this paper). Projection of
individual distance measurements from the disk's outer edge would then
require a correction of $-0.26$ $(+0.29)$ mag and $-0.16$ $(+0.17)$
mag in distance modulus for objects located at the extremes of the
disk's major and minor axes, respectively, projected behind (in front
of) the galaxy's center, compared to a face-on orientation. Accounting
for the presence or absence of a warped disk will introduce additional
systematic uncertainties: adopting a maximum extent for the warp of 3
kpc, the additional correction would be of order 0.1 mag. While
application of such corrections will largely reduce the scatter in
individual distance measurements, the uncertainties in the disk's
inclination, combined with the possibility of the presence of a warp
in the outer disk, may introduce systematic effects in excess of 0.10
mag in the resulting distance moduli.

Additional systematic uncertainties affecting the robustness of
distance determinations to the SMC relate to corrections for
reddening, absolute calibration of the relevant conversion relations,
and metallicity differences (for in-depth discussions, see also Paper
I; de Grijs 2011). Corrections for metallicity differences and
systematic offsets in the calibration relations adopted are
tracer-specific. As such, we discuss these systematic effects
separately for the Cepheid, RR Lyrae, and RC-based distances,
respectively, in Sections \ref{cepheids.sec}, \ref{rrlyrae.sec}, and
\ref{redclump.sec}. On the other hand, the effects of extinction, a
combination of absorption and scattering by dust and gas, affect all
methods to largely similar extents. Corrections for reddening are
among the most significant in the context of systematic uncertainties
feeding through into distance determinations. This is, hence, driving
development of `reddening-free’ approaches, including e.g., the
period--Wesenheit (PW) calibration relations developed for
variable-star analysis (see Section \ref{variables.sec}).

De Grijs (2011, his Chapter 6.1.1) provides a detailed discussion of
the systematic uncertainties associated with the effects of extinction
as pertaining to distance determinations. Briefly, these include
uncertainties related to our sufficiently precise knowledge of (i) the
prevailing extinction law, (ii) the intrinsic photometric properties
of one's calibration objects, and (iii) the geometry of the dust
distribution. The choice of extinction law is particularly important
when comparing similar types of objects drawn from Galactic and
Magellanic Cloud samples, since `the' Galactic extinction law (which
may, in fact, vary along different lines of sight) differs
systematically from that in the Magellanic Clouds (for recent studies,
see e.g., Dobashi et al. 2009; Bot et al. 2010; and references
therein). Nevertheless, the differences are generally $\la 0.05$ mag
at wavelengths longwards of $\lambda = 1\mu$m and shortwards of
$\lambda = 0.8 \mu$m. The significantly reduced effects of extinction
at IR wavelengths, combined with the often smaller scatter of physical
properties, is driving research efforts, e.g. in relation to
variable-star period--luminosity relations (PLRs), from the classical,
optical regime to these longer wavelengths.

Adoption of the most appropriate extinction law additionally requires
a detailed knowledge of the geometry of the mixture of dust and stars,
and the relevant filling factor, allowing for patchy versus smooth
distributions of the dust component; the commonly used `foreground
screen' geometry is often an oversimplification. For the same optical
depth, a uniform mixture of dust and stars causes less extinction than
the foreground-screen model, because part of the extinction lies
behind the source. These effects are often compounded by the unknown
effects caused by population changes, i.e., the
`age-extinction(-metallicity) degeneracy.' Finally, one has to
consider the possibility that, even if the extinction component acts
as an obscuring layer in front of the object of interest, it may not
represent a uniform layer but could be better characterized by
differential extinction. Haschke et al. (2012) provide an excellent
example of the potentially devastating effects of adopting different
assumptions for one's extinction properties. For both their Cepheid
and RR Lyrae samples in the SMC, they derive systematic uncertainties
in the resulting SMC distance modulus of 0.17--0.19 mag, depending on
whether they apply individual reddening corrections to each of their
sample objects or instead use a blanket extinction correction
pertaining to carefully selected areas. The latter assumption leads to
significantly larger distance moduli.

In this paper, we aim at extending and validating the local distance
framework established in Papers I and II by adding the SMC to our
ensemble of Local Group galaxies. As for Papers I and II, we searched
the NASA/Astrophysics Data System (ADS) article database for any
articles referring to the SMC. The volume of publications returned
from the first journal papers until the end of January 2015 included
11,095 separate entries. We systematically combed through these
papers, in reverse chronological order, looking for new or updated
distance estimates to the SMC.

We aimed at compiling a database of SMC distance determinations that
is as complete as possible for the period from January 1990 until and
including January 2015. As long as we cover a period that allows us to
discern any statistical trends, the precise choice of starting date
for our modern period is not important. For consistency with Papers I
and II, and given that all important SMC distance tracers are well
represented in the period since 1990, here we also adopt 1990 as the
start of the period of interest.\footnote{Also note that the
  individual measurements were not obtained in isolation; calibrations
  of recent data rely on calibrations of earlier results. Updates to
  the input physics are continuously implemented, thus improving the
  resulting outputs. Extending our analysis to several decades before
  the cut-off used both in this paper and in Papers I and II, would
  therefore contribute little, if anything, to the results presented
  here.} This period is covered by a total of 9746 articles in the
NASA/ADS database. We will use these for our statistical analysis. For
further reference, for the period prior to 1990, we included distance
estimates that were referred to in the body of later papers we perused
in detail: in essence, for these earlier entries we followed the
reference trail. This eventually led us to the earliest reference to
the SMC as an extragalactic object, which was in fact among the
earliest suggestions that the SMC might be an object outside of our
own Galaxy. This was proposed at a time well before the Great Debate
on the Scale of the Universe had taken place between Shapley and
Curtis, in 1920 (Curtis 1921; Shapley 1921). Indeed, Hertzsprung
(1913) boldly attempted to measure a trigonometric parallax to the
SMC, reporting a value of $10^{-4}$ arcsec. Although this corresponds
to a distance of 30,000 light years, his paper refers to a distance of
merely 3000 light years. Whether or not this was a genuine
typographical error or one of the first cases of publication bias
remains unclear (J. Lub, 2014, private communication).

\begin{figure}
\begin{center}
\includegraphics[width=\columnwidth]{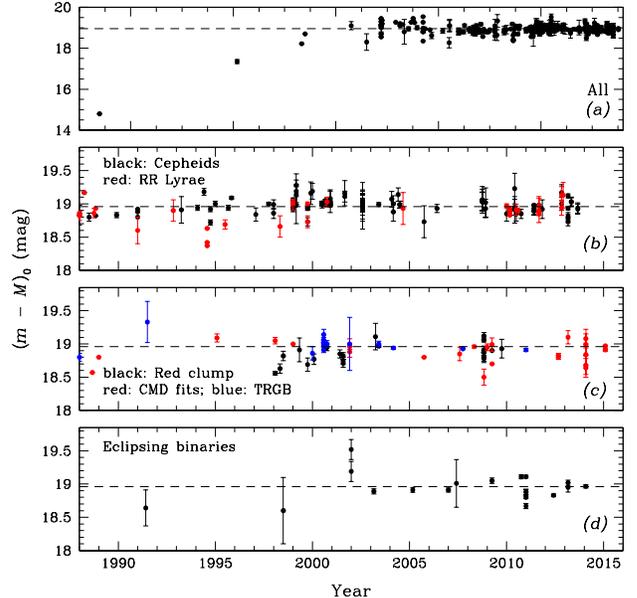}
\caption{Published extinction-corrected SMC distance moduli as a
  function of publication date (month) for all data sets pertaining to
  the SMC body or any of the galaxy's components. The horizontal
  dashed lines indicate our final, recommended distance modulus,
  $(m-M)_0 = 18.96$ mag (Section \ref{concl.sec}). (a) Full data set,
  clearly showing the historical trend. (b) Variable stars (Section
  \ref{variables.sec}). (c) Stellar population tracers (Section
  \ref{stelpop.sec}). CMD: Color--magnitude diagram. TRGB: Tip of the
  red-giant branch. (d) Eclipsing binary systems (Section
  \ref{EBs.sec}).}
\label{smcdist.fig}
\end{center}
\end{figure}

Our database analysis resulted in a total of 304 SMC distance
estimates, spanning a large range of approaches, stellar populations,
and distance tracers. Figure \ref{smcdist.fig} shows the full set of
SMC distance measurements in the database, with panels showing the
full historical data set as well as the individual distances published
since 1990 for the most commonly used tracers. As for Papers I and II,
the full database is availabe from
http://astro-expat.info/Data/pubbias.html,\footnote{For a permanent
  link to this page and its dependent pages, direct your browser
  to\\ http://web.archive.org/web/20150313155101/http://astro-expat.info/Data/pubbias.html}
as a function of both publication date and distance indicator. Its
structure is similar to that used for our LMC distances database
presented in Paper I. In the remainder of this paper, we will analyze
the distance estimates to the SMC pertaining to a number of individual
distance tracers, including eclipsing binary systems (EBs; Section
\ref{EBs.sec}), Cepheid and RR Lyrae variable stars (Section
\ref{variables.sec}), stellar population tracers such as the RC and
(red) giant stars (Section \ref{stelpop.sec}), and star clusters
(Section \ref{clusters.sec}). We will discuss the results from the
individual distance indicators and derive a common, robust mean
distance to the SMC in Section \ref{concl.sec}.

\section{Eclipsing Binary Systems}
\label{EBs.sec}

We start our analysis of the distance to the center of the SMC by
examining the galaxy's large sample of EBs, because these represent
the best geometric distance tracers at the SMC's distance. Since the
pioneering efforts by Bell et al. (1991) and Pritchard et al. (1998),
EB-based distance determination to the SMC has become fairly
routine. We carefully analyzed all relevant papers containing EB-based
distance determinations in order to base our conclusions on the most
appropriate EB sample.

First, we limited our sample to EBs associated with the SMC's main
body. This eliminated HV 2226 (Bell et al. 1991), the only SMC EB with
an individual distance estimate located in the galaxy's Wing. We next
assessed the remaining sample of 96 distance determinations for
duplicates and redundant measurements. The early distance
determinations to OGLE SMC-SC7 066175\footnote{OGLE is the Optical
  Gravitational Lensing Experiment
  (http://ogle.astrouw.edu.pl). Object OGLE SMC-SC7 066175 is located
  in OGLE SMC scan region 7.} (Pritchard et al. 1998) and OGLE SMC-SC5
202153 (Ostrov 2001) were superseded by more recent estimates by
Hilditch et al. (2005) and Harries et al. (2003), respectively, using
more up-to-date model approaches. Of the remaining objects, the sample
of Harries et al. (2003) contains four objects in common with Drechsel
\& Ne{\ss}linger (2010), while the sample of Hilditch et al. (2005)
contains four (different) EBs in common with that of North et
al. (2009, 2010; see also Gauderon et al. 2007). One SMC EB, OGLE
SMC-SC5 038089, is included in all of Harries et al. (2003), North et
al. (2010), and Drechsel \& Ne{\ss}linger (2010). However, North et
al. (2010) suggest to discard their estimate on account of unreliable
color measurements.

We checked whether there might be any systematic offsets between the
distance estimates of Harries et al. (2003) and Drechsel \&
Ne{\ss}linger (2010), and between Hilditch et al. (2005) and North et
al. (2010). Although the subsample sizes are admittedly small, we did
not find any systematic differences between either set of distance
estimates. As such, for those objects in common and which were not
affected by other deteriorating effects (see below), we adopted the
average values of Harries et al. (2003) and Drechsel \& Ne{\ss}linger
(2010) for OGLE SMC-SC5 038089, OGLE SMC-SC6 215965, OGLE SMC-SC7
243913, and OGLE SMC-SC11 030116, and those of Hilditch et al. (2005)
and North et al. (2010) for OGLE SMC-SC4 110409 and OGLE SMC-SC5
026631.

We discarded OGLE SMC-SC4 163552 from our final sample because of the
effects of a third light contribution noted by North et al. (2010). In
addition, North et al. (2010) indicated that their measurements of
OGLE SMC-SC5 180185, OGLE SMC-SC5 261267, and OGLE SMC-SC5 277080 were
affected by unreliable colors. We therefore discarded the former two
objects from our sample, given that we do not have access to
independent measurements, while for the latter object we adopted the
distance estimate of Hilditch et al. (2005).

\begin{table*}
\caption{Data set adopted to determine the best early-type EB distance modulus to the SMC.}
\label{ltEBs.tab}
\begin{center}
\tabcolsep 0.5mm
\begin{tabular}{@{}clllc|clllc@{}}
\hline \hline
Object name & \multicolumn{1}{c}{R.A. (J2000)} &
\multicolumn{1}{c}{Dec (J2000)} & \multicolumn{1}{c}{$(m-M)_0$$^b$} &
Ref.$^a$ & Object name & \multicolumn{1}{c}{R.A. (J2000)} &
\multicolumn{1}{c}{Dec (J2000)} & \multicolumn{1}{c}{$(m-M)_0$$^b$} &
Ref.$^a$ \\
(OGLE SMC) & \multicolumn{1}{c}{(hh mm ss.ss)} & 
\multicolumn{1}{c}{(dd mm ss.s)} & \multicolumn{1}{c}{(mag)} & &
(OGLE SMC) & \multicolumn{1}{c}{(hh mm ss.ss)} &
\multicolumn{1}{c}{(dd mm ss.s)} & \multicolumn{1}{c}{(mag)} \\
\hline
SC1 099121  & 00 38 51.93 & $-$73 34 33.4 & 19.29             & Hi05 & SC5 266513  & 00 50 57.34 & $-$73 12 29.4 & $19.13 \pm 0.118$ & N10  \\
SC4 056804  & 00 46 33.14 & $-$73 22 17.0 & 18.66             & Hi05 & SC5 277080  & 00 51 11.38 & $-$73 05 21.7 & 18.95             & Hi05 \\
SC4 103706  & 00 47 25.55 & $-$73 27 17.3 & 18.65             & Hi05 & {\it SC5 277080} & 00 51 11.38 & $-$73 05 21.7 & $18.52 \pm 0.056$ & N10$^e$ \\
{\it SC4 110409} &00 47 00.16 & $-$73 18 43.5 & 18.40             & Hi05 & SC5 283079  & 00 50 58.56 & $-$73 04 36.1 & $19.11 \pm 0.054$ & N10  \\
{\it SC4 110409} & 00 47 00.16 & $-$73 18 43.5 & $19.06 \pm 0.061$ & N10  & SC5 300549  & 00 51 23.57 & $-$72 52 24.1 & 18.52        & Hi05 \\
SC4 110409  & 00 47 00.16 & $-$73 18 43.5 & $18.73 \pm 0.33 $ & $^c$ & SC5 305884  & 00 51 20.17 & $-$72 49 42.9 & 18.86             & Hi05 \\
SC4 113853  & 00 47 03.95 & $-$73 15 20.5 & $19.00 \pm 0.078$ & N10  & SC5 316725  & 00 51 05.95 & $-$72 40 56.7 & 18.90             & Ha03 \\
SC4 117831  & 00 47 31.66 & $-$73 12 01.5 & $18.99 \pm 0.062$ & N10  & SC5 255984  & 00 51 29.63 & $-$73 21 38.3 & 18.54             & Hi05 \\
SC4 121084  & 00 47 32.14 & $-$73 09 08.8 & $19.28 \pm 0.051$ & N10  & SC5 311566  & 00 51 34.83 & $-$72 45 46.5 & 18.66             & Hi05 \\
SC4 121110  & 00 47 04.63 & $-$73 08 39.8 & $19.08 \pm 0.057$ & N10  & SC6 011141  & 00 52 03.95 & $-$73 18 49.1 & 19.20             & Hi05 \\
SC4 121461  & 00 47 24.66 & $-$73 09 35.1 & $19.05 \pm 0.083$ & N10  & SC6 077224  & 00 51 50.13 & $-$72 39 22.7 & 18.73             & Ha03 \\
SC4 159928  & 00 48 13.56 & $-$73 19 31.2 & $19.29 \pm 0.066$ & N10  & SC6 152981  & 00 52 41.89 & $-$72 46 22.8 & 18.64             & Hi05 \\
SC4 160094  & 00 48 10.21 & $-$73 19 37.4 & $18.96 \pm 0.102$ & N10  & SC6 158118  & 00 52 19.28 & $-$72 41 51.7 & 18.77             & Ha03 \\
{\it SC4 163552} & 00 47 53.20 & $-$73 15 57.0 & 18.49          & Hi05$^d$& SC6 180084  & 00 53 42.43 & $-$73 23 20.3 & 18.78        & Hi05 \\
{\it SC4 163552} & 00 47 53.20 & $-$73 15 57.0 & $18.35 \pm 0.079$ &N10$^d$&{\it SC6 215965} & 00 53 33.35 & $-$72 56 24.1 & 18.83   & Ha03 \\
SC4 175149  & 00 48 34.75 & $-$73 06 53.0 & $18.52 \pm 0.057$ & N10  & {\it SC6 215965} & 00 53 33.36 & $-$72 56 24.5 & $18.67 \pm 0.04 $ & DN10 \\
SC4 175333  & 00 48 15.33 & $-$73 07 05.1 & $18.61 \pm 0.074$ & N10  & SC6 215965  & 00 53 33.36 & $-$72 56 24.5 & $18.75 \pm 0.11 $ & $^e$ \\
SC5 016658  & 00 49 02.93 & $-$73 20 55.9 & $19.13 \pm 0.068$ & N10  & SC6 221543  & 00 53 39.89 & $-$72 52 19.4 & 19.09             & Hi05 \\
{\it SC5 026631} & 00 48 59.84 & $-$73 13 28.8 & 18.79             & Hi05 & SC6 251047  & 00 53 43.94 & $-$72 31 24.2 & 18.69        & Hi05 \\
{\it SC5 026631} & 00 48 59.84 & $-$73 13 28.8 & $19.13 \pm 0.036$ & N10  & SC6 311225  & 00 54 02.03 & $-$72 42 21.9 & 18.52        & Hi05 \\
SC5 026631  & 00 48 59.84 & $-$73 13 28.8 & $18.96 \pm 0.17 $ & $^c$ & SC6 319960  & 00 54 05.25 & $-$72 34 26.2 & 19.05             & Hi05 \\
SC5 032412  & 00 48 56.62 & $-$73 11 38.8 & $19.19 \pm 0.044$ & N10  & {\it SC7 066175} & 00 54 38.22 & $-$72 32 06.40& $18.6  \pm 0.5 $ & P98$^h$ \\
{\it SC5 038089} & 00 49 01.82 & $-$73 06 07.2 & $18.80 \pm 0.02 $ & DN10 & SC7 066175  & 00 54 38.22 & $-$72 32 06.4 & 18.77        & Hi05 \\
{\it SC5 038089} & 00 49 01.85 & $-$73 06 06.9 & 18.92             & Ha03 & SC7 120044  & 00 55 31.64 & $-$72 43 07.6 & 18.72        & Hi05 \\
{\it SC5 038089} & 00 49 01.85 & $-$73 06 06.9 & $18.89 \pm 0.042$ &N10$^f$ & SC7 142073  & 00 55 54.44 & $-$72 28 08.7 & 18.62      & Hi05 \\
SC5 038089  & 00 49 01.85 & $-$73 06 06.9 & $18.86 \pm 0.12 $ & $^e$ & SC7 189660  & 00 56 37.31 & $-$72 41 43.6 & 19.38             & Hi05 \\
SC5 060548  & 00 48 35.40 & $-$72 52 56.5 & 19.22             & Hi05 & SC7 193779  & 00 56 21.80 & $-$72 37 01.7 & 19.27             & Hi05 \\
SC5 095194  & 00 49 50.49 & $-$73 19 31.4 & 19.29             & Hi05 & {\it SC7 243913} & 00 56 56.34 & $-$72 49 06.4 & 19.10        & Ha03 \\
SC5 095337  & 00 49 15.34 & $-$73 22 05.8 & $19.17 \pm 0.097$ & N10  & {\it SC7 243913} & 00 56 56.34 & $-$72 49 06.4 & $19.11 \pm 0.02 $ & DN10 \\
SC5 095557  & 00 49 18.19 & $-$73 21 55.3 & $19.19 \pm 0.052$ & N10  & SC7 243913  & 00 56 56.34 & $-$72 49 06.4 & $19.11 \pm 0.1  $ & $^e$ \\
SC5 100485  & 00 49 20.02 & $-$73 17 55.5 & $18.84 \pm 0.052$ & N10  & SC7 255621  & 00 57 26.51 & $-$72 36 45.8 & 18.95             & Hi05 \\
SC5 100731  & 00 49 29.33 & $-$73 17 57.9 & $19.28 \pm 0.090$ & N10  & SC8 087175  & 00 58 30.96 & $-$72 39 14.4 & 19.10             & Hi05 \\
SC5 106039  & 00 49 20.08 & $-$73 13 35.9 & $18.95 \pm 0.050$ & N10  & SC8 104222  & 00 58 25.08 & $-$72 19 10.4 & 19.13             & Hi05 \\
SC5 111649  & 00 49 17.26 & $-$73 10 23.6 & $18.86 \pm 0.046$ & N10  & SC8 209964  & 01 00 16.02 & $-$72 12 44.3 & 18.62             & Hi05 \\
SC5 123390  & 00 49 22.61 & $-$73 03 43.3 & $18.75 \pm 0.078$ & N10  & SC9 010098  & 01 00 52.90 & $-$72 47 48.6 & 19.18             & Hi05 \\
SC5 140701  & 00 49 43.10 & $-$72 51 09.5 & 18.62             & Hi05 & SC9 047454  & 01 00 52.05 & $-$72 07 06.0 & 19.12             & Hi05 \\
SC5 180064  & 00 50 44.70 & $-$73 17 40.3 & 19.05             & Hi05 & SC9 064498  & 01 01 17.34 & $-$72 42 32.5 & 18.75             & Hi05 \\
{\it SC5 180185} & 00 50 02.71 & $-$73 17 34.2 & $19.45 \pm 0.073$ &N10$^f$&SC9 175323  & 01 03 21.27 & $-$72 05 37.8 & $18.88 \pm 0.04 $ & DN10 \\
SC5 180576  & 00 50 13.51 & $-$73 16 32.8 & $19.14 \pm 0.106$ & N10  & SC10 033878 & 01 03 21.27 & $-$72 05 37.8 & 18.84             & Ha03 \\
SC5 185408  & 00 50 24.61 & $-$73 14 55.8 & $19.12 \pm 0.057$ & N10  & SC10 037156 & 01 03 28.82 & $-$72 01 28.9 & 19.11             & Hi05 \\
SC5 202153  & 00 50 27.93 & $-$73 03 16.1 & 19.13             & Ha03 & SC10 094559 & 01 05 06.82 & $-$72 24 57.4 & 18.64             & Hi05 \\
{\it SC5 202153} & 00 50 27.95 & $-$73 03 16.5 & $19.36 \pm 0.22 $ &O01$^g$&SC10 108086 & 01 05 30.57 & $-$72 01 21.4 & 18.76        & Hi05 \\
SC5 208049  & 00 50 44.98 & $-$72 58 44.5 & 19.48             & Hi05 & SC10 110440 & 01 05 09.59 & $-$71 58 42.3 & 18.29             & Hi05 \\
SC5 243188  & 00 51 18.78 & $-$73 30 16.3 & 19.33             & Hi05 & {\it SC11 030116} & 01 06 24.86 & $-$72 12 48.3 & 18.71       & Ha03 \\
{\it SC5 261267} & 00 51 35.04 & $-$73 17 11.1 & $19.35 \pm 0.068$ &N10$^f$&{\it SC11 030116} & 01 06 24.88 & $-$72 12 48.7 & $18.88 \pm 0.04 $ & DN10 \\
SC5 265970  & 00 51 28.12 & $-$73 15 17.9 & $19.25 \pm 0.048$ & N10  & SC11 030116 & 01 06 24.88 & $-$72 12 48.7 & $18.79 \pm 0.11 $ & $^e$ \\
SC5 266015  & 00 51 16.73 & $-$73 13 02.7 & $19.23 \pm 0.038$ & N10  & SC11 057855 & 01 07 31.44 & $-$72 19 52.9 & 18.92             & Ha03 \\
SC5 266131  & 00 51 35.63 & $-$73 12 44.1 & $19.11 \pm 0.081$ & N10  & {\it HV 2226}        & 01 24       & $-$73.3       & $18.64 \pm 0.27 $ & B91$^i$ \\
\hline \hline
\end{tabular}
\end{center}
\flushleft
Objects referenced using italic font were not included in our final
analysis for reasons indicated in these footnotes.\\
$^a$ References: B91 -- Bell et al. (1991); DN10 -- Drechsel \&
Ne{\ss}linger (2010); Ha03 -- Harries et al. (2003); Hi05 -- Hilditch
et al. (2005); N10 -- North et al. (2010); O01 -- Ostrov 2001; P98 --
Pritchard et al. (1998).\\
$^b$ Where no uncertainties are provided by the original authors, we
adopted uncertainties of 0.10 mag to determine the weighted mean. \\
$^c$ Average of Hi05 and N10; the error indicates the range. \\
$^d$ Discarded because of a third light contribution. \\
$^e$ Average of Ha03 and DN10. \\
$^f$ Discarded because of unreliable colors. \\
$^g$ Superseded by Ha03. \\
$^h$ Superseded by Hi05. \\
$^i$ Wing EB.
\end{table*}

These considerations left us with a final SMC EB sample of 75
objects. The full data set at the basis of this analysis is provided
in Table \ref{ltEBs.tab}. All of these SMC EBs were composed of
early-type (O- and B-type) components. The geometric average position
of all sample objects is located firmly within the SMC's main body, at
RA (J2000) = 00$^{\rm h}$ 52$^{\rm m}$ 58.2$^{\rm s}$, Dec (J2000) =
$-72^\circ$ 55$'$ 14.7$''$, i.e., slightly south of the main body's
stellar density center, RA (J2000) = 00$^{\rm h}$ 52$^{\rm m}$
44.8$^{\rm s}$, Dec (J2000) = $-72^\circ$ 49$'$ 43$''$ listed in the
NASA/IPAC Extragalactic
Database.\footnote{http://ned.ipac.caltech.edu}

Although neither Harries et al. (2003) nor Hilditch et al. (2005)
include uncertainties on their individual distance measurements, the
former authors suggest that their typical systematic and random
uncertainties are of order 0.10 and 0.15 mag, respectively, in
distance modulus. The latter authors refer to Harries et al. (2003) to
support their claim of a 0.10 mag systematic uncertainty. For the
purpose of determining a weighted mean distance modulus, we adopt
uncertainties of 0.10 mag in distance modulus for those EBs without
individual uncertainty estimates. Increasing this to 0.15 mag does not
appreciably change our result. The resulting weighted mean distance
modulus to our sample of 75 early-type EBs (etEBs) is
\begin{equation}
(m-M)_0^{\rm etEB} = 18.93 \pm 0.03 \mbox{ mag.}
\end{equation}
Adopting a normal distribution for the distance estimates leads to
$(m-M)_0^{\rm etEB} = 18.95$ mag and a standard deviation (Gaussian
$\sigma$) of 0.26 mag. This compares well with the mean distance
modulus quoted by Harries et al. (2003), $(m-M)_0 = 18.89 \pm 0.04 \pm
0.10$ mag, where the first and second uncertainty estimates represent
the statistical and systematic errors, respectively. Similarly,
Hilditch et al. (2005) find $\langle (m-M)_0 \rangle = 18.91 \pm 0.03
\pm 0.1$ mag for their full EB sample.

Since distance estimates to etEBs are subject to fairly large
systematic uncertainties owing to the need for adoption of stellar
atmosphere models (cf. Pietrzy\'nski et al. 2013; Paper I), which are
notoriously difficult to correct for, longer-period late-type (cool)
giant EBs are preferable as geometric distance tracers. Unfortunately,
the numbers of such SMC EBs with reliable distance estimates are still
small. Nevertheless, Graczyk et al. (2014) combined new measurements
of four late-type EBs (ltEBs) with their earlier estimate of the
distance to OGLE-SC10 137844 (Graczyk et al. 2012, 2013) to arrive at
\begin{equation}
(m-M)_0^{\rm ltEB} = 18.965 \pm 0.025 \pm 0.048 \mbox{ mag},
\end{equation}
where the uncertainties again refer to the statistical and systematic
errors, respectively.

\section{Variable Stars as Distance Indicators}
\label{variables.sec}

In the absence of significant numbers of geometric distance tracers,
variable-star PLRs and PW relations have become fundamental tools to
study the nearest rungs of the astrophysical distance ladder, although
lingering systematic uncertainties persist. The most commonly used
PLRs are derived from Cepheid and RR Lyrae variable stars, which we
will cover separately in this section. In addition, the SMC hosts Mira
and semi-regular variables, red-giant-branch (RGB) pulsators, and
carbon stars. All of these tracers have been used in attempts to
determine the galaxy's distance; we will refer to these efforts in our
discussion of giant stars as distance tracers in Section
\ref{trgb.sec}.

\subsection{Cepheids}
\label{cepheids.sec}

Cepheids are the most commonly used distance tracers in relation to
the SMC. Since records began (Shapley 1940), we have collected some
120 individual Cepheid-based distance measurements to the SMC or its
components. In this section, we will explore what we can learn from
the roughly 70 modern, post-2000 measurements included in our
database.

The majority of Cepheid-based SMC distance estimates rely on
classical, fundamental-mode (FU) Cepheid PLRs and PW relations, while
a small number of additional measurements are based on first- and
second-overtone (FO/SO) pulsators, as well as on double- or mixed-mode
(`beat'), Type II, and bump Cepheids. The numbers of these latter
measurements are too small to perform a proper statistical analysis,
with the possible exception of the FO Cepheid-based
distances. However, their associated distance moduli can be used to
corroborate the statistical analysis facilitated by the much larger
number of classical FU Cepheids. We will now first explore what we can
learn from this most common type of Cepheids.

Despite their common use as distance tracers, significant systematic
uncertainties remain in the application of Cepheid light-curve
observations to the distance problem. First, metallicity differences
between comparison populations may affect the resulting PLR slopes
significantly (e.g., Sakai et al. 2004; Tammann et al. 2008; Bono et
al. 2010; Matsunaga et al. 2011; Groenewegen 2013), although these
effects are reduced at near-IR wavelengths (e.g., Storm et al. 2000;
Bono et al. 2010) and they seem absent at the longer, mid-IR
wavelengths probed by the {\sl Spitzer Space Telescope} (Majaess et
al. 2013). In contrast, the reddening-free Wesenheit magnitudes do not
appear to depend on a population's metallicity, even at optical
wavelengths (cf. Inno et al. 2013a).

Second, PLR-based distance calibrations are most commonly done in a
relative sense, by deriving the differential distance modulus between
a calibration population's PLR and that of the target sample. The
majority of Cepheid-based distance estimates to the SMC use Galactic
Cepheids as their baseline for absolute distance determination. A
number of authors have pointed out that at least two types of
systematic uncertainties may affect the validity of such an
approach. Most subtly, Galactic Cepheid PLRs at optical and near-IR
wavelengths are linear for all periods, within the intrinsic
uncertainties. In contrast, the LMC PLRs are known to exhibit a clear
`break' (a change of slope) in the relations at a period of
approximately 10 days; it appears that the SMC PLRs may exhibit either
a break at $\log(P/{\rm d}) \simeq 0.4$ or a downward curvature
towards shorter periods (e.g., Tammann et al. 2008; Bono et al. 2010;
Matsunaga et al. 2011). This would clearly invalidate any direct
differential distance modulus determination, yet many authors proceed
along these lines nevertheless. Once again, it turns out that use of
the reddening-free PW relations avoids this critical issue: Inno et
al. (2013a) use a sample of 2571 FU Cepheids observed through
$JHK_{\rm s}$ filters to conclude that the PW slopes in both the
Magellanic Clouds and the Milky Way are linear. Still, in a follow-up
paper Inno et al. (2013b) take great care to determine the
differential LMC--SMC distance modulus only at a pivotal period of
$\log(P/{\rm d}) = 0.5$ (0.3) for FU (FO) Cepheids.

Second, absolute distance calibration based on comparison with
Galactic objects is known to be plagued by significant systematic
effects. These calibrations are often based on parallax measurements,
which are unfortunately very small and have typical uncertainties in
excess of 30\%. {\sl Hubble Space Telescope}-based parallaxes are less
seriously affected by these parallax errors (e.g., Benedict et
al. 2002, 2007) than the earlier {\sl Hipparcos} measurements,
although the revised {\sl Hipparcos} parallaxes (van Leeuwen 2007)
provide a significantly improved calibration data
set.\footnote{Udalski (2000) eloquently explained and clearly showed
  that absolute magnitude calibration of bright Galactic Cepheids,
  based on the original {\sl Hipparcos} parallaxes, should be
  approached with significant caution. Such calibrations predict
  $M_V^{\rm Ceph} = -4.2$ mag for $\log(P/{\rm d}) = 1.0$ (Feast \&
  Catchpole 1997; Lanoix et al. 1999; Groenewegen \& \hbox{Oudmaijer}
  2000), which Udalski (2000) assesses as too bright (see also
  Abrahamyan 2004), i.e., they are affected by large systematic
  uncertainties in the zero-point calibration (for a discussion, see
  Paper I). Calibrations of fainter objects tend to be more reliable;
  they are usually based on `quasi-geometrical' methods such as the
  Barnes--Evans variant of the Baade--Wesselink (BW) surface
  brightness method (e.g., Storm et al. 2000, 2004, 2011; Barnes et
  al. 2004; Groenewegen 2013) or pre-{\sl Hipparcos} Galactic
  calibrations (e.g., Laney \& Stobie 1994), which are less affected
  by systematic uncertainties in the photometric zero point owing to,
  e.g., uncertain extinction corrections or Lutz--Kelker-type biases
  (cf. de Grijs 2011; his Chapter 6.1.2). {\sl Hubble Space Telescope}
  and revised {\sl Hipparcos} parallax calibrations are, fortunately,
  used fairly extensively to determine distances to Local Group
  galaxies (for SMC distances, see e.g., Majaess et al. 2008; Inno et
  al. 2013a).}

In the interest of full disclosure, we point out that in this paper we
have opted to use differential LMC--SMC distance moduli where
provided, combined with $(m-M)_0^{\rm LMC} = 18.50$ mag (cf. Paper I),
to compile a homogenized database of SMC distances. This is
particularly important in the context of pre-2000 LMC/SMC distance
determinations (e.g., Udalski 2000), compared with later measurements
based on significantly overlapping tracer samples (e.g., Inno et
al. 2013a), given the persistent `long' versus `short' LMC distance
dichotomy that affected this field prior to the new millennium (for a
detailed discussion, see Paper I). After all, establishing a robust
LMC distance modulus was one of the main aims of Paper I; we are now
using that result to our advantage in this paper.

When we keep in mind these caveats and combine these concerns with the
complex geometry of the SMC, it is indeed highly surprising that the
post-2000 (and, in fact, many earlier) Cepheid-based SMC distance
determinations cluster very closely around an SMC distance modulus of
$(m-M)_0 \sim 19.0$ mag. At first glance, this might imply that (i)
none of these effects are sufficiently important to have a significant
effect, (ii) multiple caveats may affect many of these distance
determinations simultaneously, somehow counteracting each others'
effects, or (iii) we are witnessing the effects of (presumably
unconscious) publication bias.

For our detailed analysis of the body of FU Cepheid-based SMC
distances, we will first homogenize the distance scale by `correcting'
when necessary any distance estimate to the commonly adopted LMC
benchmark distance modulus of $(m-M)_0^{\rm LMC} = 18.50$ mag (for a
detailed discussion, see Paper I). Many of the authors cited in our
database provide, in fact, differential LMC--SMC distance moduli. In
such cases, we homogenize the database by adding these differential
values to the canonical LMC distance modulus adopted here (for similar
approaches applied to a large number of nearby galaxies, see also
Ferrarese et al. 2000; Sakai et al. 2004).

Of the PLR- and PW-based post-2000 SMC distance estimates in our
database, we now highlight a few that need special care in our
subsequent statistical analysis. Udalski (2000) derives an LMC
distance of $(m-M)_0 = 18.24$ mag based on the $I$-band PW relation,
although his differential LMC--SMC distance modulus is $\Delta \mu_0 =
0.51 \pm 0.05$, which is well within the commonly accepted
range. McCumber et al. (2005) selected a subset of Cepheids from the
$BV$-based compilation of Mathewson et al. (1986) to determine a
distance modulus of $(m-M)_0 = 18.73 \pm 0.24$ mag to a field in the
SMC's Wing. Since we are interested in deriving the most appropriate
distance to the {\it main body} of the SMC, we will discard this
measurement. In addition, careful analysis of the premises on which
Mathewson et al. (1986) based their Cepheid distance scale implies
that they adopted $(m-M)_0^{\rm LMC} = 18.45$ mag ($D_{\rm LMC} = 49$
kpc). Finally, Haschke et al. (2012) attempted to correct their
optical ($VI$), OGLE-based sample of 2522 FU Cepheids for the effects
of foreground extinction both by taking the area-averaged attenuation
and by comparing the observed and intrinsic colors of their sample
stars (i.e., essentially using a period--luminosity--color relation)
to derive individual extinction values. The resulting difference in
the derived distance modulus is $\Delta (m-M)_0 = 0.17$--0.19
mag. This thus serves as a strong indication of the importance of
minimizing the systematic uncertainties. We will henceforth use their
distance modulus resulting from extinction correction on a
star-by-star basis.

In addition to the PLR- and PW-based Cepheid distance estimates to the
SMC, more direct attempts have been made---using much smaller sample
sizes---by application of the BW/Barnes--Evans `quasi-geometrical'
surface brightness approach (e.g., Storm et al. 2000, 2004, 2011;
Barnes et al. 2004; Groenewegen 2013). Based on only five Cepheids,
Storm et al. (2000) find an SMC distance modulus of $(m-M)_0 = 19.19
\pm 0.12$ mag if using $V$ and $K$ magnitudes, but $(m-M)_0 = 18.90
\pm 0.07$ mag based on $VR_{\rm J}$ photometry. Using the same five
stars, they derive $(m-M)_0 = 18.88 \pm 0.14$ mag (corrected for depth
effects) and $(m-M)_0 = 18.92 \pm 0.14$ mag in their 2004 and 2011
papers, respectively, using the same near-IR surface brightness
technique.

Groenewegen (2013) uses six SMC Cepheids and a number of different
near-IR BW-type approaches to estimate $(m-M)_0 = 18.73$--18.81
mag. He acknowledges that these distances are smaller than expected
from the full body of SMC distance measurements and states that this
``is not predicted by theoretical investigations, but these same
investigations do not predict a steep dependence on period [as found
  here] either, indicating that additional theoretical work is
warranted.''  One possible solution to reconcile these shorter BW
distances with the longer PLR/PW-based distances may be found in the
possible metallicity dependence of the $p$ (`projection')
factor\footnote{Projection ($p$) factors are commonly used to convert
  radial to pulsation velocities.} used in BW analyses (Groenewegen
2013).

In view of these concerns, we will not include BW-type analyses in
determining the weighted mean SMC distance. We are therefore left with
an ensemble of 32 SMC distance estimates published between June 2000
and August 2013 based on PLR and PW analyses. The resulting weighted
mean distance, using the statistical uncertainties as weights, is
\begin{equation}
(m-M)_0^{\rm FU\,Ceph} = 19.00 \pm 0.02 \mbox{ mag},
\end{equation}
with a standard deviation (implying both a significant line-of-sight
depth and the lingering effects of systematic uncertainties) of 0.08
mag. Our weighting approach is fully justified on statistical grounds,
because the minimum numbers of Cepheids contributing to the individual
SMC distance measurements are 91 (Ferrarese et al. 2000) and 94 (Sakai
et al. 2004).\footnote{The single exception to this statement is the
  sample composed of 13 FU Cepheids used by McCumber et al. (2005),
  although we do not use this measurement. The latter authors state
  that ``the distribution of [these] 13 Cepheids ...  appears to be
  roughly Gaussian,'' while their quoted uncertainty of 0.24 mag
  reflects the more uncertain nature of their mean distance modulus
  compared with the other recent measurements included in our
  database.} As such, our method is not compromised by finite sampling
properties, in which case the distributions of individual FU Cepheid
distance moduli would formally follow a $t$ distribution with a given
number of degrees of freedom. However, provided that the number of
data points used to derive the respective distance moduli is larger
than approximately 50, the results from a $t$ distribution are almost
the same as those from a normal (Gaussian) distribution, and the
associated error bar reflects the most likely uncertainty in the
mean. Our working samples thus represent large populations where the
effects of small-sample statistics can be ignored. Since the limit of
a $t$ distribution for a large number of data points is a normal
(Gaussian) distribution, the statistical inferences for a large number
of data points will be the same. Hence, this fully justifies our
approach. The remaining 29 distance moduli are based on much larger
numbers of FU Cepheids, often in excess of a few thousand
objects.\footnote{Groenewegen (2000): 2048 (OGLE)/1511 ({\sl
    Hipparcos}); Udalski (2000): up to 3300; Pietrzy\'nski et
  al. (2003): No numbers quoted, but the sample is based on the large
  OGLE database; Majaess et al. (2008): 2140; Feast (2011) and
  Matsunaga et al. (2011): 2436; Haschke et al. (2012): 2522; Inno et
  al. (2013a,b): 2571 and 2626, respectively.} The only exceptions
here include the samples used by Abrahamyan (2004), Bono et al. (2008,
2010), and Majaess et al. (2013), which contain 234, $\sim$200, 344,
and $\sim$100 objects (based on inspection of their Fig. 1),
respectively. Nevertheless, these numbers of FU Cepheids still meet
our minimum rule-of-thumb number of 50 quoted above.

We will now place these results for FU Cepheids in the context of
other Cepheid-based distances. Our database includes 13 SMC distance
estimates based on FO Cepheids, taken from four articles by the same
group (Bono et al. 2001, 2002; Inno et al. 2013a,b). These authors
used $I$-band and near-IR PW analysis to obtain their results. Their
weighted mean distance is $(m-M)_0^{\rm FO\,Ceph} = 19.01 \pm 0.02$
mag. An interesting result from Inno et al. (2013a) is that the PLR
slope of the FO pulsators differs from the corresponding slope of
their FU counterparts. This means that the FO Cepheids should not be
`fundamentalized' to improve the statistical sample of FU pulsators,
as is often done in the literature. We note, however, that Groenewegen
(2000) analyzed their FU and FO Cepheids separately and deemed them
sufficiently consistent (contrary to the result of Inno et al. 2013a)
to lead to a single differential LMC--SMC distance modulus.

The single SO Cepheid distance modulus reported is $(m-M)_0^{\rm
  SO\,Ceph} = 19.11 \pm 0.08$ mag (Bono et al. 2001), while
double-mode (or beat) Cepheid OGLE ($VI$)-based analysis implies
$(m-M)_0^{\rm beat} = 19.05 \pm 0.02$ mag (Kov\'acs 2000). All of
these measurements are internally consistent and commensurate with the
weighted mean distance we derived for the FO Cepheids. Indeed, Inno et
al. (2013a) similarly concluded that mixed-mode Cepheids follow the
same PW relations in the SMC as their FO counterparts.

Finally, we also retrieved SMC distance moduli based on both Type II
and bump Cepheids. The single bump-Cepheid measurement yields
$(m-M)_0^{\rm bump\,Ceph} = 18.93 \pm 0.06$ mag (Keller \& Wood 2006),
where we determined the uncertainty ourselves based on the published
data, because the authors only provided the uncertainty on the mean
rather than the standard deviation of the distribution (see the
relevant discussion in Paper I). We have found four independent SMC
distance estimates based on Type II Cepheids (Majaess et al. 2009;
Ciechanowska et al. 2010; Matsunaga et al. 2011), resulting in a
weighted mean distance of $(m-M)_0^{\rm TII} = 18.87 \pm 0.06$ mag,
for which we used the random uncertainties as weights. These objects
are closer in nature to RR Lyrae stars (see Section \ref{rrlyrae.sec})
than to classical Cepheids (e.g., Bono et al. 1997; Wallerstein 2002).

\begin{table*}
\caption{Adopted, homogenized Cepheid distances used in this paper.}
\label{cepheids.tab}
\begin{center}
\tabcolsep 0.5mm
\begin{tabular}{@{}cclccl@{}}
\hline \hline
Publ. date & $(m-M)_0$ & \multicolumn{1}{c}{Reference} & PLR/ & Filter(s) & \multicolumn{1}{c}{Notes} \\
(mm/yyyy) & (mag) & & PW \\
\hline
\multicolumn{6}{c}{\bf FU Cepheids}\\
\hline
06/2000 & $18.997 \pm 0.024$ & Ferrarese et al. (2000)   & PLR & $J$   & Madore \& Freedman (1991) calibration \\
06/2000 & $19.013 \pm 0.022$ & Ferrarese et al. (2000)   & PLR & $H$   & Madore \& Freedman (1991) calibration \\
06/2000 & $18.989 \pm 0.022$ & Ferrarese et al. (2000)   & PLR & $K$   & Madore \& Freedman (1991) calibration \\
06/2000 & $18.99  \pm 0.05 $ & Ferrarese et al. (2000)   & PLR & $JHK$ & Cardelli et al. (1989) reddening law \\
09/2000 & $19.01  \pm 0.05 $ & Udalski (2000)            & PW  & $I$   \\
11/2000 & $19.08  \pm 0.11 $ & Groenewegen (2000)        & PW  & $VI$  \\
11/2000 & $19.04  \pm 0.17 $ & Groenewegen (2000)        & PLR & $K_{\rm s}$ \\
11/2000 & $19.01  \pm 0.03 $ & Groenewegen (2000)        & PW  & $VI$  & Udalski et al. (1999) calibration \\
11/2000 & $18.975 \pm 0.022$ & Groenewegen (2000)        &     &       & Mean of 6 IR determinations \\
11/2000 & $19.004 \pm 0.015$ & Groenewegen (2000)        &     &       & Mean of all 8 determinations \\
05/2003 & $18.99  \pm 0.03 $ & Pietrzy\'nski et al. (2003)&PLR & $K$   \\
01/2004 & $19.070 \pm 0.119$ & Abrahamyan (2004)         & PLR & $BVIJHK$ \\
06/2004 & $18.99  \pm 0.05 $ & Sakai et al. (2004)       & PLR & $VI$  & Madore \& Freedman (1991) calibration \\
06/2004 & $18.99  \pm 0.05 $ & Sakai et al. (2004)       & PLR & $VI$  & Udalski et al. (1999) calibration \\
09/2005 & $18.78  \pm 0.24 $ & McCumber et al. (2005)    & PLR & $I_V$ & Subset of Mathewson et al. (1986) \\
09/2008 & $19.06  \pm 0.20 $ & Bono et al. (2008)        & PW  & $BV$  & $P > 6$ days, metallicity corrected \\
09/2008 & $19.03  \pm 0.14 $ & Bono et al. (2008)        & PW  & $VI$  & $P > 6$ days \\
09/2008 & $19.04  \pm 0.14 $ & Bono et al. (2008)        & PW  & $BI$  & $P > 6$ days \\
09/2008 & $19.06  \pm 0.16 $ & Bono et al. (2008)        & PW  & $BVI$ & $P > 6$ days \\
11/2008 & $18.93  \pm 0.14 $ & Majaess et al. (2008)     & PLR & $VI$  \\
11/2008 & $19.02  \pm 0.22 $ & Majaess et al. (2008)     & PLR & $VJ$  \\
05/2010 & $19.23  \pm 0.23 $ & Bono et al. (2010)        & PW  & $BV$  \\
05/2010 & $18.95  \pm 0.12 $ & Bono et al. (2010)        & PW  & $VI$  \\
05/2010 & $18.91  \pm 0.20 $ & Bono et al. (2010)        & PW  & $JK_{\rm s}$ \\
05/2011 & $18.98  \pm 0.01 $ & Matsunaga et al. (2011)   & PLR,PW & $IK$  \\
05/2011 & $18.93  \pm 0.05 $ & Matsunaga et al. (2011)   & PLR,PW & $IK$  & Metallicity corrections \\
08/2011 & $18.98  \pm 0.01 $ & Feast (2011)              & PW  & $VI$  \\
10/2012 & $19.00  \pm 0.10 $ & Haschke et al. (2012)     & PLR &       & Individual reddening \\
02/2013 & $18.93  \pm 0.02 $ & Inno et al. (2013a)       & PW  & $VIJHK_{\rm s}$ & Systematic uncertainty 0.10 mag \\
04/2013 & $19.03  \pm 0.06 $ & Inno et al. (2013b)       & PW  & $VIJHK_{\rm s}$ \\
08/2013 & $18.938 \pm 0.077$ & Majaess et al. (2013)     & PLR & $3.6 \mu$m \\
08/2013 & $18.921 \pm 0.075$ & Majaess et al. (2013)     & PLR & $4.5 \mu$m \\
\hline
\multicolumn{6}{c}{\bf FO Cepheids}\\
\hline
08/2001 & $19.16  \pm 0.19 $ & Bono et al. (2001)        & PW  & $VI$  & Incl. FO components in FO/SO pulsators \\
07/2002 & $19.06  \pm 0.13 $ & Bono et al. (2002)        & PW  & $IK$  & Theoretical calibration \\
07/2002 & $18.98  \pm 0.21 $ & Bono et al. (2002)        & PLR & $I$   & Theoretical calibration \\
07/2002 & $19.02  \pm 0.19 $ & Bono et al. (2002)        & PLR & $K$   & Theoretical calibration \\
07/2002 & $19.02  \pm 0.14 $ & Bono et al. (2002)        & PW  & $IK$  & Empirical calibration \\
07/2002 & $18.91  \pm 0.17 $ & Bono et al. (2002)        & PLR & $I$   & Empirical calibration \\
07/2002 & $18.97  \pm 0.35 $ & Bono et al. (2002)        & PLR & $K$   & Empirical calibration \\
07/2002 & $19.04  \pm 0.09 $ & Bono et al. (2002)        &     & $I$   & Weighted mean, theoretical calibration \\
07/2002 & $18.98  \pm 0.10 $ & Bono et al. (2002)        &     & $I$   & Weighted mean, empirical calibration \\
07/2002 & $19.04  \pm 0.11 $ & Bono et al. (2002)        &     & $K$   & Weighted mean, theoretical calibration \\
07/2002 & $19.01  \pm 0.13 $ & Bono et al. (2002)        &     & $K$   & Weighted mean, empirical calibration \\
02/2013 & $19.12  \pm 0.03 $ & Inno et al. (2013a)       & PW  & $VIJHK_{\rm s}$ & Systematic uncertainty 0.10 mag \\
04/2013 & $19.03  \pm 0.07 $ & Inno et al. (2013b)       & PW  & $VIJHK_{\rm s}$ \\
\hline
\multicolumn{6}{c}{\bf Type II Cepheids}\\
\hline
12/2009 & $18.85  \pm 0.11 $ & Majaess et al. (2009)     & PW  & $VI$ \\
09/2010 & $18.85  \pm 0.07 $ & Ciechanowska et al. (2010)& PLR & $JK$ & Systematic uncertainty 0.07 mag \\
05/2011 & $18.90  \pm 0.07 $ & Matsunaga et al. (2011)   & PW  & $IK$ \\
05/2011 & $18.89  \pm 0.05 $ & Matsunaga et al. (2011)   & PLR & $I$  \\
\hline
\multicolumn{6}{c}{\bf Other Cepheid Types}\\
\hline
08/2000 & $19.05  \pm 0.017$ & Kov\'acs (2000)           &     & $VI$  & Double-mode (beat) Cepheids \\
        &                    &                           &     &       & Syst. unc. 0.043 mag; theoretical models \\
08/2001 & $19.11  \pm 0.08 $ & Bono et al. (2001)        & PW  & $VI$  & SO Cepheids\\
05/2006 & $18.93  \pm 0.06 $ & Keller \& Wood (2006)     &     & $VR$  & Bump Cepheids; Pulsation models \\
\hline \hline
\end{tabular}
\end{center}
\end{table*}

\begin{figure}
\begin{center}
\includegraphics[width=\columnwidth]{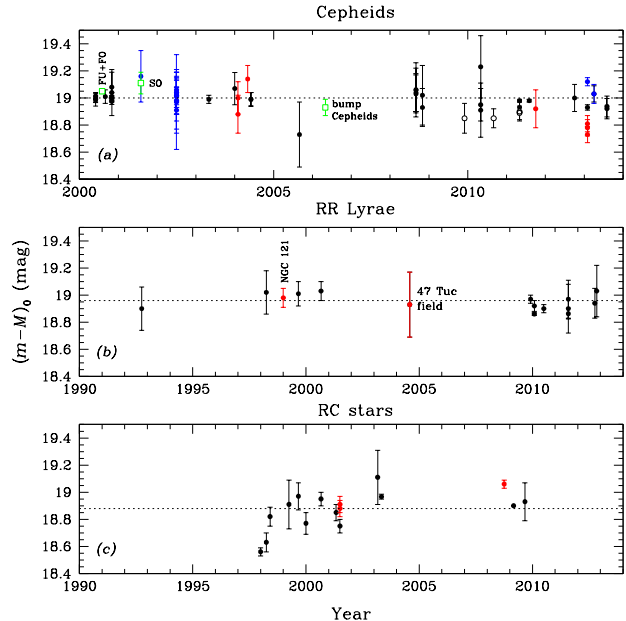}
\caption{Cleaned data sets used for our statistical SMC distance
  analysis. The horizontal dotted lines represent the weighted mean
  levels pertaining to each sample of specific distance indicators
  (see text). (a) Cepheid samples. Black solid bullets: FU Cepheids,
  PLR/PW-based distances. Red solid bullets: FU Cepheids, BW-type
  distance estimates. Blue solid bullets: FO Cepheids. Black open
  circles: Type II Cepheids. (b) RR Lyrae samples. Red data points
  represent distances to NGC 121 and an SMC background field behind
  the Galactic globular cluster 47 Tucanae. (c) RC stars. Red data
  points represent distance estimates to NGC 121.}
\label{fig2.fig}
\end{center}
\end{figure}

Table \ref{cepheids.tab} lists all Cepheid-based distances used in
this paper, for all different Cepheid types considered, recalibrated
to a canonical LMC distance modulus of $(m-M)_0^{\rm LMC} = 18.50$ mag
where necessary. Figure \ref{fig2.fig}a provides a summary of our
final Cepheid data sets. The black solid bullets represent PLR- and
PW-based distance estimates using samples of FU Cepheids, while the
red solid bullets indicate BW-type distances to FU Cepheids. Blue
solid bullets correspond to FO Cepheid-based distances, and black open
circles indicate distance estimates based on Type II
Cepheids. Finally, the green open squares are labeled with the
specific types of Cepheids used for their determination.

\subsection{RR Lyrae stars}
\label{rrlyrae.sec}

RR Lyrae stars are the most numerous variable stars in old stellar
populations. Various attempts have been made to determine the distance
to the SMC using its population of RR Lyrae variables. Here we will
focus on the main achievements since 1990; our full database includes
RR Lyrae-based distance estimates since the early attempts by
Thackeray \& Wesselink (1953, 1954) of distance determination to NGC
121, the oldest globular cluster in the SMC. In this section, we will
only consider RR Lyrae-based SMC distance estimates pertaining to the
galaxy's field population in the main body. We specifically exclude
any efforts made at determining the distance to NGC 121, but we will
return to that object in Section \ref{ngc121.sec}.

This restriction leaves us with 22 individual field-star distance
estimates published in 17 different articles between October 1992 and
November 2012. Of these, four publications (Udalski 1998a; Udalski et
al. 1999; Kapakos et al. 2011; Kapakos \& Hatzidimitriou 2012) use as
their basis LMC distance moduli that deviate from our adopted
canonical value, $(m-M)_0^{\rm LMC} = 18.50$ mag. We either applied
the relevant corrections to the SMC distance moduli reported by these
authors or used the differential LMC--SMC distance modulus (if given),
combined with $(m-M)_0^{\rm LMC} = 18.50$ mag, to arrive at a
homogenized set of SMC distance estimates.

We carefully explored the assumptions underlying the different
analyses pertaining to RR Lyrae-based distance estimates to the
SMC. With increasing numbers of RR Lyrae observations in the SMC
becoming available over the period of interest, the mean SMC distance
modulus appears to have reached a stable value. Nevertheless, we
caution that some of the analyses used in this section must be handled
carefully.

For instance, Sandage et al. (1999) based their distance determination
of $(m-M)_0 = 19.00 \pm 0.03$ mag on very careful calibration, but
they adopted two questionable assumptions. First, their observational
sample consisted of RR Lyrae from both Walker \& Mack (1988) and Smith
et al. (1992). The former comprised a sample of RR Lyrae stars in NGC
121, while the latter consisted of field stars in the vicinity of both
NGC 121 and NGC 361. Udalski (1998b) suggests that NGC 121 is located
$0.08 \pm 0.04$ mag behind the SMC's center (cf. Section
\ref{ngc121.sec}), so that this assumption could introduce a
systematic offset in the derived distance to the SMC. Second, the
metallicity--luminosity relation they adopted is characterized by a
very steep slope---$M_V \propto 0.30$ [Fe/H]---which is outside the
range commonly agreed upon (for a discussion, see Paper I). Both
assumptions might, in fact, conspire to lead to an `SMC distance
modulus' that falls inside the range implied by other studies, but the
basic premise of the approach used in this case is questionable. For
these reasons, we will not include this result in our analysis.

Despite significant improvements in our understanding of the
degeneracies affecting RR Lyrae-based distance calibration, including
those owing to the effects of metallicity differences, the effects of
extinction remain troublesome. This can be seen clearly by considering
the set of distance moduli provided by Haschke et al. (2012), who
attempted to correct their OGLE-based sample of 1494 fundamental-mode
RR Lyrae (RRab) for the effects of foreground extinction both by
taking the area-averaged attenuation and by comparing the observed and
intrinsic colors of their sample RR Lyrae stars to derive individual
extinction values. The resulting difference in the derived distance
modulus (applied to the same RR Lyrae sample) is $\Delta (m-M)_0
\simeq 0.2$ mag. Following our approach for the Cepheid-based
distances, we will therefore use their distance modulus resulting from
extinction correction on a star-by-star basis.

Extinction effects may also have caused a systematic overestimate of
the SMC distance by Kapakos \& Hatzidimitriou (2012). These authors
identify a systematic difference in their reddening corrections
compared with their earlier work (Kapakos et al. 2011), by $\Delta
E(B-V) = 0.02 \pm 0.05$ mag to $\Delta E(B-V) = 0.08 \pm 0.02$ mag for
the SMC's inner and outer regions, respectively, thus systematically
reducing the distance difference between the LMC and SMC by $\Delta
(m-M)_0 = 0.07 \pm 0.17$ mag to $\Delta (m-M)_0 = 0.27 \pm 0.07$ mag.
Additional, although likely small, systematic uncertainties pertain to
the calibration approaches adopted, with more recent analyses using a
Fourier light-curve decomposition method (Jurcsik \& Kov\'acs 1996) to
derive metallicities (e.g., Deb \& Singh 2010; Kapakos et al. 2010,
2011; Kapakos \& Hatzidimitriou 2012), followed by absolute-magnitude
calibration based on either theoretical models (e.g., Weldrake et
al. 2004) or robust observational analysis (e.g., Clementini et
al. 2003). The latter authors explored a range of different
$M_V$--[Fe/H] luminosity--metallicity relations to calibrate the
absolute magnitudes of the 77 RRab, 38 RRc, and 10 RRd variables in
their sample of LMC RR Lyrae stars, as well as cross-calibrations with
other distance indicators. Their recommended calibration relation is
characterized by a slope of $\Delta M_V({\rm RR})/\Delta$[Fe/H] $=
0.214 \pm 0.047$, which is consistent with the concensus value
(cf. Paper I). The LMC zero point resulting from their RR Lyrae
analysis, combined with BW calibration and the statistical parallax
method, is consistent with the `short' distance scale, although use of
RC stars and contemporary reddening estimates move their LMC distance
modulus closer to the canonical value recommended in Paper I.
Nevertheless, and in view of the systematic uncertainties affecting
these calibration relations, we have opted to use relative LMC--SMC
distance moduli where provided, combined with $(m-M)_0^{\rm LMC} =
18.50$ mag, since relative distance moduli are significantly less
affected by lingering systematic effects.

\begin{table*}
\caption{Homogenized field RR Lyrae distances considered in this paper.}
\label{rrlyrae.tab}
\begin{center}
\begin{tabular}{@{}ccll@{}}
\hline \hline
Publ. date & $(m-M)_0$$^a$ & \multicolumn{1}{c}{Reference} & \multicolumn{1}{c}{Notes} \\
(mm/yyyy) & (mag) \\
\hline
10/1992 & $18.90  \pm 0.16$ & Smith et al. (1992)    & Field near NGC 361 \\
04/1998 & $18.66  \pm 0.16$ & Udalski (1998a)        \\
00/1999 & $19.02  \pm 0.05$ & Reid (1999)            & Based on Udalski (1998a) \\
09/1999 & ${\it 19.00  \pm 0.03}$ & Sandage et al. (1999)  & Metallicity corrected \\
09/1999 & $19.01  \pm 0.09$ & Udalski et al. (1999)  \\
09/2000 & $19.03  \pm 0.07$ & Udalski (2000)         \\
08/2004 & $18.93  \pm 0.24$ & Weldrake et al. (2004) & 47 Tuc field \\
12/2009 & $18.97  \pm 0.03$ & Szewczyk et al. (2009) & Systematic uncertainty 0.12 mag \\
02/2010 & $18.86  \pm 0.01$ & Deb \& Singh (2010)    & RRab, mean \\
02/2010 & ${\it 18.83  \pm 0.01}$ & Deb \& Singh (2010)    & RRab, intensity-weighted mean \\
02/2010 & ${\it 18.84  \pm 0.01}$ & Deb \& Singh (2010)    & RRab, phase-weighted mean \\
02/2010 & $18.92  \pm 0.04$ & Deb \& Singh (2010)    & RRc, mean \\
02/2010 & ${\it 18.89  \pm 0.04}$ & Deb \& Singh (2010)    & RRc, intensity-weighted mean \\
02/2010 & ${\it 18.89  \pm 0.04}$ & Deb \& Singh (2010)    & RRc, phase-weighted mean \\
06/2010 & $18.91  \pm 0.08$ & Majaess (2010)         \\
07/2010 & $18.90  \pm 0.03$ & Kapakos et al. (2010)  \\
08/2011 & $18.90  \pm 0.18$ & Kapakos et al. (2011)  & RRab \\
08/2011 & $18.97  \pm 0.14$ & Kapakos et al. (2011)  & RRc \\
08/2011 & $18.863 \pm 0.04$ & Feast (2011)           & $K$, corrected for metallicity effects \\
10/2012 & ${\it 19.13  \pm 0.13}$ & Haschke et al. (2012)  & Area-averaged reddening \\
10/2012 & $18.94  \pm 0.11$ & Haschke et al. (2012)  & Individual reddening \\
11/2012 & $19.11  \pm 0.19$ & Kapakos \& Hatzidimitriou (2012) \\
\hline \hline
\end{tabular}
\end{center}
\flushleft
$^a$ Distance moduli rendered in italic font were not used for the
determination of the weighted mean RR Lyrae-based SMC distance
derivation in this paper, for reasons discussed in the text.
\end{table*}

Finally, except where specifically indicated in our database, most
authors base their RR Lyrae distance estimates on RRab-type stars. A
small number of authors (Smith et al. 1992; Weldrake et al. 2004;
Szewczyk et al. 2009) base their results on a mixture of RRab and
first-overtone pulsators (RRc stars), where they fundamentalize the
RRc stars. To calculate the weighted mean distance implied by field RR
Lyrae in the SMC, we combine RRab and RRc-based distance estimates and
their statistical uncertainties; we include only the `mean' distances
given by Deb \& Singh (2010; see the database notes for details),
which thus leaves us with a final sample of 16 distance
measurements. Table \ref{rrlyrae.tab} provides an overview of our
homogenized RR Lyrae-based SMC distances data set published between
1990 and 2015, adopting the canonical LMC distance modulus,
$(m-M)_0^{\rm LMC} = 18.50$ mag where relevant. The distance moduli
highlighted in italic font were not used for the determination of the
weighted mean field RR Lyrae-based SMC distance. The latter is
\begin{equation}
(m-M)_0^{\rm RR} = 18.96 \pm 0.02 \mbox{ mag},
\end{equation}
with a standard deviation of 0.06 mag.

Figure \ref{fig2.fig}b provides a summary of our final RR Lyrae data
set. We have indicated the distances resulting from analysis of the
NGC 121 RR Lyrae (Reid 1999) as well as those in the 47 Tuc field
(Weldrake et al. 2004) separately, using red data points. Since these
measurements relate to specific observational fields, they do not
necessarily accurately reflect the distance to the SMC's center. We
will discuss distance determinations to the SMC's star clusters in
detail in Section \ref{clusters.sec}.

\section{Stellar Population Tracers}
\label{stelpop.sec}

Next, we will discuss the distance determinations resulting from
careful analysis of well-defined stellar population features along the
red-giant branch, including the magnitude of the RC and the tip of the
red-giant branch (TRGB).

\subsection{The Red Clump as a Standard Candle}
\label{redclump.sec}

RC stars are the low- to intermediate-mass ($\sim$0.7--$2 M_\odot$,
depending on chemical composition) analogs of the helium-burning
horizontal-branch stars typically seen in old globular
clusters. Theoretical models imply that their absolute luminosity
depends only weakly or even negligibly on age and metallicity,
particularly at wavelengths longward of the $I$ band (Paczy\'nski \&
Stanek 1998: $I$; Alves 2000; Grocholski \& Sarajedini 2002; Alves et
al. 2002; Sarajedini et al. 2002: $J,K$; for discussions, see
Pietrzy\'nski et al. 2010; de Grijs 2011, his Chapter 3.2.2). For
instance, Stanek \& Garnavich (1998) found an $I$-band variance of the
RC's absolute magnitude of only $\sim 0.15$ mag. In the context of SMC
distance measurements, Udalski (1998a) established that the absolute
$I$-band RC magnitude in SMC star clusters is virtually independent of
age for ages between 2 Gyr and 10 Gyr. Although RC stars span a larger
age range in many stellar populations, this result implies that the RC
magnitude is useful as a standard candle to a large fraction of the
SMC's stellar population. Similarly, Grocholski \& Sarajedini (2002)
showed that for ages between $\sim$2 Gyr and 6 Gyr and $-0.5 \le {\rm
  [Fe/H]} \le 0$ dex, the intrinsic variation in the RC's absolute
$K$-band magnitude is minimized (see also Alves 2000; but see Salaris
\& Girardi 2005 and Groenewegen 2008 for discussions of population
corrections).

Indeed, the lack of any age dependence for these ages is not subject
to debate. In the context of our database of SMC distance estimates,
which span many decades, the slope of any metallicity dependence is,
however. The latter is quoted as 0.19 and 0.21 mag dex$^{-1}$ (in
[Fe/H]) by Popowski (2000) and Cole (1998), respectively, while
Udalski (1998a) advocates 0.09 mag dex$^{-1}$. Depending on a
population's mean metallicity (and its uniformity), compared with that
in the solar neighborhood where the {\sl Hipparcos} calibration of the
RC clump magnitude, $M_I^0 = -0.23 \pm 0.03$ mag (Stanek \& Garnavich
1998), is usually taken as the baseline, this difference may lead to
{\it systematic differences} in SMC RC magnitude of up to 0.1 mag for
typical SMC star cluster metallicities ranging from [Fe/H] = $-0.7$
dex to [Fe/H] = $-1.5$ dex (Udalski 1998a).

Keeping this discussion in mind, we set off on a careful analysis of
the RC-based distance measurements to the SMC contained in our
database; see Table \ref{redclump.tab} for the RC data set considered
in this paper. Early efforts were led by Udalski and his collaborators
(Udalski 1998a,b, 2000; Udalski et al. 1998) based on both SMC OGLE
field regions (particularly scans of the low-density fields SC1, SC2,
SC10, and SC11 in the outer galaxy) and the galaxy's star cluster
population. At the time of these publications, the debate regarding a
possible metallicity dependence of the RC's $I$-band magnitude was
particularly heated, resulting in continuous updates of the SMC's
distance modulus from a low value of $(m-M)_0 = 18.65 \pm 0.03 \mbox{
  (statistical)} \pm 0.06$ (systematic) mag (Udalski 1998a) to a high
value of $(m-M)_0 = 18.95 \pm 0.05$ mag (Udalski 2000). Intermediate
and higher values, corresponding to a stronger metallicity dependence,
were also favored by other contemporary authors (Cole 1998; Twarog et
al. 1999; Popowski 2000), leading to a robust theoretical
determination of $(m-M)_0 = 18.85 \pm 0.06$ mag by Girardi \& Salaris
(2001).

More recent determinations of RC-based distances have focused on
distances to its star clusters. Although we will discuss the distance
estimates resulting from star cluster analysis separately, here we
specifically address relevant results based on their RC
magnitudes. Crowl et al. (2001) explored the RC's use as a distance
indicator to 12 SMC clusters, although only a subset (NGC 152, NGC
361, NGC 411, NGC 416, and Kron 28) are actually associated with the
galaxy's main body (for a clear overview, see their Fig. 1). The RC
magnitudes of these five clusters combined, with their statistical
uncertainties used as weights, lead to a weighted mean SMC distance
modulus of $(m-M)_0 = 18.88 \pm 0.15$ ($18.71 \pm 0.11$) mag, adopting
Burstein \& Heiles (1982) (Schlegel et al. 1998) foreground extinction
estimates. We will return to the use of star cluster samples in
Section \ref{clusters.sec}.

Of the more recent determinations, Alcaino et al. (2003), Glatt et
al. (2008c), and Cignoni et al. (2009) base their distance estimates
on star clusters located well outside the SMC's main body, with the
exception of NGC 416 studied by Glatt et al. (2008c), for which they
determine $(m-M)_0 = 18.90 \pm 0.07$ mag. Pietrzy\'nski et al. (2003)
and Sabbi et al. (2009) focus on field regions and find, respectively,
$(m-M)_0 = 18.93 \pm 0.03$ mag (which is affected by lingering
systematic uncertainties in the $K$ band) versus $(m-M)_0 = 18.89 \pm
0.45$ mag (field SFH1) and $(m-M)_0 = 18.97 \pm 0.27$ mag (field
SFH4). The latter fits are calibrated using the Bertelli et al. (1994)
isochrones. This is identical to the SMC distance estimate based on
field RC stars of Udalski et al. (1999), $(m-M)_0 = 18.97 \pm 0.09$
mag, recalibrated for a Large Magellanic Cloud (LMC) distance modulus
of $(m-M)_0^{\rm LMC} = 18.50$ mag.

\begin{table*}
  \caption{Homogenized RC-based SMC distances.}
\label{redclump.tab}
\begin{center}
\begin{tabular}{@{}ccll@{}}
\hline \hline
Publ. date & $(m-M)_0$$^a$ & \multicolumn{1}{c}{Reference} & \multicolumn{1}{c}{Notes} \\
(mm/yyyy) & (mag) \\
\hline
01/1998 & ${\it 18.56  \pm 0.03}$ & Udalski et al. (1998)$^b$ & Systematic uncertainty 0.06 mag \\
04/1998 & ${\it 18.63  \pm 0.07}$ & Uldalski (1998a)$^b$   \\
06/1998 & $18.82  \pm 0.07$  & Cole (1998)                 & Systematic uncertainty 0.13 mag \\
09/1998 & ${\it 18.65  \pm 0.08}$ & Udalski (1998b)$^b$    & Clusters \\
04/1999 & $18.91^{+0.18}_{-0.16}$ & Twarog et al. (1999)   \\
09/1999 & $18.97  \pm 0.10$  & Udalski et al. (1999)       \\
01/2000 & $18.77  \pm 0.08$  & Popowski (2000)             \\
09/2000 & $18.95  \pm 0.05$  & Udalski (2000)              \\
05/2001 & $18.85  \pm 0.06$  & Girardi \& Salaris (2001)   \\
07/2001 & $18.71  \pm 0.06$  & Crowl et al. (2001)         & 5 clusters,$^c$ Schlegel et al. (1998) reddening \\
07/2001 & $18.82  \pm 0.05$  & Crowl et al. (2001)         & 5 clusters,$^c$ Burstein \& Heiles (1982) reddening \\
03/2003 & ${\it 19.11 \pm 0.2}$ & Alcaino et al. (2003)    & NGC 458 (cluster located far to the NE of the SMC body) \\
05/2003 & $18.967 \pm 0.018$ & Pietrzy\'nski et al. (2003) & $K$ \\
10/2008 & $18.90 \pm 0.07$   & Glatt et al. (2008b)        & NGC 416 (main-body cluster) \\
03/2009 & ${\it 18.9}$       & Cignoni et al. (2009)       & NGC 602 (Wing cluster) \\
09/2009 & $18.89  \pm 0.45$  & Sabbi et al. (2009)         & Field SFH1 \\
09/2009 & $18.97  \pm 0.27$  & Sabbi et al. (2009)         & Field SFH4 \\
\hline \hline
\end{tabular}
\end{center}
\flushleft
$^a$ Distance moduli rendered in italic font were not used for the
determination of the weighted mean RC-based SMC distance
derivation in this paper (see text, Notes, and these footnotes).\\
$^b$ The earlier values published by Udalski's team were ignored given
that they continuously updated their numbers based on improved input
physics. \\
$^c$ Based on 5 clusters associated with the SMC's main body, i.e.,
NGC 152, NGC 361, NGC 411, NGC 416, and Kron 28.
\end{table*}

At the start of the period of interest considered here, the dependence
of absolute RC magnitudes on ages and metallicities was not well
understood and resulted in short distances to the Magellanic
Clouds. Recently, Groenewegen (2008) provided updated $I$- and
$K$-band calibrations in the Two-Micron All-Sky Survey (2MASS)
photometric system, $\langle M_I^{\rm RC} \rangle = -0.22 \pm 0.03$
mag and $\langle M_K^{\rm RC} \rangle = -1.54 \pm 0.04$ mag. The
latter is somewhat fainter than Grocholski \& Sarajedini's (2002)
calibration, $\langle M_K^{\rm RC} \rangle = -1.61 \pm 0.04$ mag,
which Groenewegen (2008) attributed to the need to apply population
corrections, caused by selection effects affecting the calibration
reference stars. The weighted mean value resulting from combining all
measurements pertaining to the SMC's main body (or objects associated
with it) published since 1998, again adopting the individual
uncertainties as weights, yields
\begin{equation}
(m-M)_0^{\rm RC} = 18.88 \pm 0.03 \mbox{ mag},
\end{equation}
with a standard deviation of 0.08 mag.

Figure \ref{fig2.fig}c provides an overview of our final RC data
set. We have indicated the distances resulting from analyses of the
NGC 121 RC stars (Crowl et al. 2001; Glatt et al. 2008a) separately,
using red data points. However, note that these measurements are
surrounded by some controversy. Crowl et al. (2001) used the RC
determination in this cluster from Mighell et al. (1998), based on
these latter authors' assumption that NGC 121 is an intermediate-age
cluster. If, on the other hand, NGC 121 is indeed much older (as we
argue in Section \ref{ngc121.sec}, given that it hosts $> 10$ Gyr-old
RR Lyrae stars), the cluster's `RC' stars are more likely red
horizontal-branch stars, which are not known to be good standard
candles. Nevertheless, Glatt et al. (2008a) argue convincingly that
NGC 121 is a few billion years younger than the canonical globular
cluster age in the Milky Way, so that the object may indeed be a
transition-type cluster.

\subsection{Giant Stars}
\label{trgb.sec}

Among red-giants-based distance determinations, those based on the
TRGB, the maximum absolute luminosity reached by first-ascent red
giants with ages in excess of $\sim 1$--2 Gyr, are most commonly
used. The TRGB marks the onset of helium fusion in their
electron-degenerate helium cores. Its absolute bolometric magnitude
varies by only 0.1 mag for a wide range of metallicities and ages
(Iben \& Renzini 1983; Da Costa \& Armandroff 1990; Salaris \& Cassisi
1997; Madore et al. 2009). Although the TRGB's $I$-band magnitude has
become firmly established as a local distance indicator, there is a
systematic offset of 0.1 mag between the TRGB and Cepheid distance
scales (Tammann et al. 2008). At the same time, the metallicity
dependence of the Cepheid PLR has been calibrated using the TRGB
method (Rizzi et al. 2007; see also Sanna et al. 2008), which implies
a worrying degree of circular reasoning.

Only few TRGB-based distance determinations have been published for
the SMC's main body. Cioni et al. (2000) used a combination of
near-infrared $JHK_{\rm s}$ passbands and observations from the
DENIS\footnote{Deep Near Infrared Survey of the Southern Sky;
  http://cds.u-strasbg.fr/denis.html} database to derive $(m-M)_0 =
19.02 \pm 0.04$ mag. Udalski (2000) and Pietrzy\'nski et al. (2003)
used the $I$-band TRGB magnitude and OGLE observations to derive a
very similar distance to the bulk of the SMC stars,
\begin{equation}
(m-M)_0^{\rm TRGB} = 19.00 \pm 0.04 \mbox{ mag},
\end{equation}
while these authors quote additional systematic uncertainties $\Delta
(m-M)_0 = 0.07$ mag (Udalski 2000), resulting from reddening and
calibration errors, to $\Delta (m-M)_0 \simeq 0.20$ mag (Pietrzy\'nski
et al. 2003), caused by calibration differences between the $I$ and
$K$ bands. These TRGB-based distances are somewhat larger, at the
1--2$\sigma$ level, than those resulting from other giant-star-based
tracers, including Mira and semi-regular variable stars (Kiss \&
Bedding 2004), carbon stars (Soszy\'nski et al. 2007), and RGB
pulsators (Tabur et al. 2010).

\section{Star Clusters}
\label{clusters.sec}

Many of the SMC's populous star clusters are located well away from
the galaxy's main body. In Section \ref{redclump.sec} we specifically
highlighted a small number of clusters that we consider firmly
associated with the bulk of the SMC's stellar
population. Nevertheless, by considering the galaxy's entire star
cluster population, we may gain additional insights into the most
appropriate distance to its center of mass. In addition, since star
clusters often contain multiple distance tracers simultaneously,
distances to individual star clusters can be used to cross-validate
individual methods of distance determination and identify possible
systematic offsets. This is our main aim in assessing the various
distance determinations to NGC 121 available in the
literature. Subsequently, we will combine the individual measurements
available for a sample of SMC clusters to obtain a mean distance to
the galaxy's gravitational center.

\subsection{NGC 121}
\label{ngc121.sec}

NGC 121 is the oldest populous star cluster in the SMC; its
metallicity is the lowest among the SMC's cluster population, [Fe/H] =
$-1.71 \pm 0.10$ dex (e.g., Udalski 1998b; Crowl et al. 2001). It has
been studied extensively and is host to a number of different distance
tracers. This makes the cluster a suitable testbed for our assessment
of the importance of any systematic effects among the latter. Table
\ref{ngc121.tab} includes the `best' distance moduli to NGC 121, based
on our perusal of the relevant literature.

\begin{table}
\caption{`Best' distance measures to the old SMC cluster NGC 121.}
\label{ngc121.tab}
\begin{center}
\begin{tabular}{@{}ccl@{}}
\hline \hline
$(m-M)_0$ (mag)  & Tracer   & \multicolumn{1}{c}{Reference} \\
\hline
$18.98 \pm 0.07$ & RR Lyrae & Reid (1999) \\
$18.88 \pm 0.06$ & RC$^a$   & Crowl et al. (2001) \\
$18.91 \pm 0.06$ & RC$^b$   & Crowl et al. (2001) \\
$19.0  \pm 0.4 $ & TRGB     & Dolphin et al. (2001) \\
$18.98 \pm 0.10$ & HB level & Dolphin et al. (2001) \\
$18.96 \pm 0.04$ & CMD fit  & Dolphin et al. (2001) \\
$18.96 \pm 0.02$ & CMD fit  & Glatt et al. (2008a) \\
$19.06 \pm 0.03$ & RC       & Glatt et al. (2008a) \\
\hline \hline
\end{tabular}
\end{center}
\flushleft
$^a$ Burstein \& Heiles (1982) extinction adopted\\
$^b$ Schlegel et al. (1998) extinction adopted
\end{table}

Most tracers tend towards a larger distance modulus of around $(m-M)_0
\sim 19.0$ mag, although a spread of order 0.1 mag is clearly
implied. We only have access to multiple measurements for distance
determinations to the cluster based on the use of RR Lyrae stars and
fits to its CMD. The difference between the RR Lyrae-based distances
suggested by Nemec et al. (1994) and that of Reid (1999) is
sufficiently large so as to warrant a detailed examination. Nemec et
al. (1994) included two such distance estimates, based on two
different sets of photometric measurements. They point out, following
Walker \& Mack (1988), that the earlier photographic-plate
measurements of four RR Lyrae stars by Graham (1975) are
systematically fainter by $\sim 0.2$ mag in both the $B$ and $V$ bands
than the CCD-based observations of Walker \& Mack (1988). For the SMC
field, Nemec et al. (1994) could only compare their estimate with that
of Graham (1975). The distance modulus to NGC~121 based on $B$-band
photometry reported by Walker \& Mack (1988) is, to within their
mutual $1\sigma$ photometric uncertainties of 0.03 mag, fully
consistent with the SMC field distance of Graham (1975). Udalski
(1998b) also concluded from a comparison with newly obtained CCD data
that the Graham (1975) SMC field photometry is in excellent agreement
with their new measurements. In view of these considerations, and
given the difficulty of deriving accurate photographic photometry in
crowded fields which most likely affected Graham's (1975) cluster
photometry, Reid (1999) used the Walker \& Mack (1988) cluster
photometry to base his distance determination on.

The main physical difference between the Nemec et al. (1994) and Reid
(1999) distance estimates to NGC 121 is found in their use of
calibration object. Nemec et al. (1994) base their distance
calibration on the distance to the Galactic globular cluster M15,
while Reid (1999) reports a differential distance modulus with respect
to the LMC. Reid (1999) quotes a mean $V$-band magnitude of LMC
cluster RR Lyrae of $\langle V_0 \rangle = 18.98$ mag (Walker 1994),
which he compared with the equivalent value for NGC 121 RR Lyrae in
the SMC, $\langle V_0 \rangle = 19.46 \pm 0.07$ mag (Walker \& Mack
1988), to derive $\Delta \mu_0 = 0.48 \pm 0.07$ mag (irrespective of
any calibration relations adopted), and hence $(m-M)_0^{\rm SMC} =
18.98 \pm 0.07$ mag. A careful assessment of the choices made by Nemec
et al. (1994) implies that their distance calibration corresponds to
an LMC distance modulus of $(m-M)_0^{\rm LMC} = 18.35$ mag. Correcting
the resulting distance modulus to the canonical LMC distance modulus
results in an updated distance to NGC 121 of $(m-M)_0^{\rm NGC\,121} =
18.78$ mag (no individual uncertainties quoted, although the authors
provide an upper limit of 0.2 mag). For consistency with Papers I and
II, as well as with the choices made in this paper, we adopt Reid's
(1999) estimate.

The differences in NGC 121 distance moduli based on stellar population
tracers are generally less than 0.10 mag, with the exception of the
distance estimates based on measurements of the RC magnitude (Crowl et
al. 2001 versus Glatt et al. 2008a; but see the caveat mentioned in
Section \ref{redclump.sec}). Crowl et al. (2001) applied a correction
of $+0.093$ mag to the theoretical $M_V({\rm RC})$ values of Girardi
et al. (2000; see also Girardi \& Salaris 2001). This explains the
systematic difference between the Crowl et al. (2001) and Glatt et
al. (2008a) NGC 121 distance moduli. Since we adopted the Girardi et
al. RC calibration in Section \ref{redclump.sec}, for reasons of
internal consistency we will adopt the Glatt et al. (2008a) RC-based
distance.

Based on these considerations, the resulting weighted mean distance
modulus to NGC 121 is $(m-M)_0 = 18.98 \pm 0.02$ mag. In terms of
differential distance moduli, Udalski (1998b) suggests that NGC 121 is
located $0.08 \pm 0.04$ mag behind the SMC's center, thus leading to
\begin{equation}
(m-M)_0^{\rm NGC\,121\rightarrow SMC} = 18.90 \pm 0.04 \mbox{ mag}.
\end{equation}

\subsection{The SMC Cluster Population}
\label{ensembles.sec}

Finally, it is instructive to determine the mean distance to the SMC's
cluster population. Although most of the galaxy's star clusters are
located well outside its main body, their mean distance gives us
additional insights into the relevant distance scale. Among our
database of SMC distance determinations, three groups of studies
provide homogeneous sets of cluster distances (Crowl et al. 2001;
Glatt et al. 2008a,b,c; Dias et al. 2014). We strongly prefer to use a
homogeneous baseline for our analysis of the ensemble of SMC clusters.

\begin{table}
\caption{SMC clusters with published distance estimates.}
\label{clusters.tab}
\begin{center}
\tabcolsep 0.5mm
\begin{tabular}{@{}lccl@{}}
\hline \hline
Cluster & $(m-M)_0$ & Tracer   & \multicolumn{1}{c}{Reference$^a$} \\
        & (mag) \\
\hline
47 Tuc field & $18.93 \pm 0.24$ & RR Lyrae & Weldrake et al. (2004) \\
AM 3         & $18.99 \pm 0.16$ & CMD fits & Dias et al. (2014) \\
BS 90        & $18.85 \pm 0.1 $ & CMD fits & Rochau et al. (2007) \\
BS 196       & $18.95 \pm 0.05$ & CMD fits & Bica et al. (2008) \\
HW 1         & $18.84 \pm 0.16$ & CMD fits & Dias et al. (2014) \\
HW 40        & $19.08 \pm 0.14$ & CMD fits & Dias et al. (2014) \\
ICA 16       & $19.05 \pm 0.05$ & CMD fits & Demers \& Battinelli (1998) \\
Kron 3       & $18.80 \pm 0.05$ & RC       & Weighted average (C01, G08) \\
Kron 28      & $18.78 \pm 0.09$ & RC       & Weighted average (C01) \\
Kron 44      & $18.92 \pm 0.04$ & RC       & Weighted average (C01) \\
Lindsay 1    & $18.67 \pm 0.06$ & RC       & Weighted average (C01, G08) \\
Lindsay 2    & $18.68 \pm 0.14$ & CMD fits & Dias et al. (2014) \\
Lindsay 3    & $18.64 \pm 0.14$ & CMD fits & Dias et al. (2014) \\
Lindsay 38   & $19.03 \pm 0.04$ & RC       & Weighted average (C01, G08) \\
Lindsay 113  & $18.47 \pm 0.07$ & RC       & Weighted average (C01) \\
NGC 121      & $18.98 \pm 0.02$ & Multiple & This paper \\
NGC 152      & $18.96 \pm 0.19$ & RC       & Weighted average (C01) \\
NGC 330      & $18.82 \pm 0.03$ & Cepheids & Weighted average$^b$ \\
NGC 339      & $18.78 \pm 0.02$ & RC       & Weighted average (C01, G08) \\
NGC 361      & $18.61 \pm 0.12$ & RC       & Weighted average (C01) \\
NGC 411      & $18.57 \pm 0.14$ & RC       & Weighted average (C01) \\
NGC 416      & $18.89 \pm 0.07$ & RC       & Weighted average (C01, G08) \\
NGC 419      & $18.50 \pm 0.12$ & RC       & Glatt et al. (2008b) \\
NGC 602-A    & $18.7$           & RC       & Cignoni et al. (2009) \\
Unnamed      & $18.8$           & CMD fits & McCumber et al. (2005) \\
cluster \\
\hline \hline
\end{tabular}
\end{center}
\flushleft 
$^a$ C01: Crowl et al. (2001); G08: Glatt et al. (2008c);\\ 
$^b$ Weighted average of two distance estimates based on different
stellar models, one without and one with moderate overshooting (Sebo
\& Wood 1994).
\end{table}

Table \ref{clusters.tab} includes the full set of 25 SMC clusters for
which individual distance determinations are available in the
literature, based on a variety of tracers. We first checked for
duplicates in distance estimates to these sample objects. (Note that
we did not include NGC 121 in this analysis, give that we addressed
the distance to this cluster in the previous section.) Most
importantly, Crowl et al. (2001) and Glatt et al. (2008c) have five
clusters in common, i.e., NGC 339, NGC 416, Lindsay 1, Lindsay 38, and
Kron 3. From the Crowl et al. (2001) results, we considered their
distance estimates using both the Burstein \& Heiles (1982) and the
Schlegel et al. (1998) estimates of the foreground extinction.  These
authors offer the choice of adopting either a constant absolute RC
magnitude, independent of age or metallicity, or adoption of the
assumption that the absolute RC magitude is a function of both age and
metallicity. We adopted the latter assumption and used the individual
cluster distance moduli as tabulated by Crowl et al. (2001). As
pointed out by these authors, the distances based on the Schlegel et
al. (1998) extinction estimates are systematically shorter; a
comparison between the Glatt et al. (2008c) distances and the Crowl et
al. (2001) values using Burstein \& Heiles (1982) extinction estimates
shows that the Crowl et al. (2001) distances tend to be somewhat
shorter than those of Glatt et al. (2008c), which is owing to the
different RC-magnitude calibrations adopted by the two different
teams. In other words, this is a systematic effect. Since one of our
aims is to understand the systematic uncertainties involved in Local
Group distance determinations, we decided to calculate weighted
average values for those clusters with both Crowl et al. (2001) and
Glatt et al. (2008c) distances, without any pre-selection.

Three additional clusters were found to have duplicate distance
determinations. The distance to AM 3 was determined by both Da Costa
(1999) and also recently by Dias et al. (2014). We chose to retain the
latter value because of its inclusion in the homogeneous set of
distance determinations of Dias et al. (2014), although the more
recent estimate is essentially identical to the earlier
determination. Second, Dias et al. (2014) published an updated
distance estimate for Lindsay 2, which supersedes their earlier,
significantly larger value from Dias et al. (2008). We selected the
more recent determination for further analysis. (In addition, their
earlier value originated from a conference contribution while the more
recent distance was published in a peer-reviewed article.) Finally,
NGC 361 was analyzed by both Smith et al. (1992) and Crowl et
al. (2001). For reasons of homegeneity, we opted to use the Crowl et
al. (2001) values, while we also noted that the Smith et al. (1992)
distance related to a field near the cluster rather than to NGC 361
itself.

The projected geometric mean center position of the entire star
cluster system thus selected is found at RA (J2000) = 00$^{\rm h}$
52$^{\rm m}$ 41$^{\rm s}$, Dec (J2000) = $-72^\circ 40' 28''$, which
coincides with a location in the densest stellar region of the SMC's
main body. Note that we did not set out to select an unbiased cluster
sample, although we also point out that our final sample of 25
clusters is not necessarily biased in any way in relation to the
resulting set of distances. The weighted mean distance to this
arbitrary set of 25 SMC clusters is
\begin{equation}
(m-M)_0^{\rm clusters} = 18.81 \pm 0.03 \mbox{ mag},
\end{equation}
with a standard deviation of 0.17 mag. The latter value includes both
depth effects and systematic uncertainties.

This mean distance compares well with previous determinations of the
star cluster centroid distance, although based on smaller numbers of
clusters. Crowl et al.'s (2001) RC-based distance determinations to
their small sample of 12 clusters led to a mean distance of $(m-M)_0 =
18.82 \pm 0.05$ mag ($18.71 \pm 0.05$ mag) assuming Burstein \& Heiles
(1982) (Schlegel et al. 1998) foreground extinction, while the average
RC-based distance to the six clusters studied by Glatt et al. (2008c)
was found at $(m-M)_0 = 18.87 \pm 0.03$ mag.

\section{Final Recommendation}
\label{concl.sec}

In an effort to provide a firm mean distance estimate to the SMC, and
thus place it within the internally consistent Local Group distance
framework we established recently, we performed extensive analysis of
the published literature to compile the largest database available to
date containing SMC distance estimates.

We highlight the need for such an effort by pointing out that almost
all authors who derive either `short' or `long'
distances\footnote{Note that the terms `short' and `long' used in this
  context simply refer to the extremes of the published SMC distance
  range, and not to the `short' and `long' distance scales used to
  refer to LMC distances in previous decades (cf. Paper I).} to the
SMC based on their chosen distance indicator and tracer sample
selectively refer to a subset of recent (and not-so-recent) distance
estimates that support their result. The danger of this habit
persisting in the literature is that one loses sight of the global
picture. We aim at remedying this situation by providing estimates of
the mean SMC distance based on a large number of distance tracers.
Table \ref{summary.tab} offers a summary of the mean distances
determined in this paper.

\begin{table}
\caption{Mean distances to the SMC based on a range of distance indicators.}
\label{summary.tab}
\begin{center}
\begin{tabular}{@{}ccl@{}}
\hline \hline
$(m-M)_0$ (mag)  & $\sigma$ (mag) & \multicolumn{1}{c}{Tracer} \\
\hline
$18.93 \pm 0.03$ & 0.26 & Early-type EBs \\
$18.965\pm0.025$ & 0.048$^a$ & Late-type EBs \\
$19.00 \pm 0.02$ & 0.08 & FU Cepheids \\
$19.01 \pm 0.02$ & 0.06 & FO Cepheids \\
$18.96 \pm 0.02$ & 0.06 & RR Lyrae \\
$18.88 \pm 0.03$ & 0.08 & RC stars \\
$19.00 \pm 0.04$ & --   & TRGB \\
($18.90 \pm 0.04$& --   & NGC 121) \\
($18.81 \pm 0.03$& 0.17 & Star clusters) \\
\hline \hline
\end{tabular}
\end{center}
\flushleft
$^a$ The standard deviation given for the late-type EBs is the
systematic uncertainty reported by Graczyk et al. (2014).
\end{table}

Throughout the paper, we have emphasized the important role attributed
to systematic uncertainties. In addition to the corrections for
geometric and depth effects required because of the galaxy's complex
nature, we also pointed out lingering systematic uncertainties in the
absolute distance calibrations using a variety of stellar tracers, as
well as those owing to uncertain extinction corrections. Nevertheless,
if we take the simple weighted mean of the distances given in Table
\ref{summary.tab}, except for the bracketed values pertaining to the
SMC's star clusters, we obtain our final recommendation for the `mean'
SMC distance,
\begin{equation}
(m-M)_0^{\rm SMC} = 18.96 \pm 0.02 \mbox{ mag}.
\end{equation}
This value is fully consistent with the recommendation by Graczyk et
al. (2014) based on their analysis of both early- and late-type EBs,
the distance indicator thought to be least affected by systematic
uncertainties owing to poorly understood physics. It is indeed
encouraging to note that the most recent SMC distance determination,
which is based on mid-IR PLR analysis of FU Cepheids, also yields
$(m-M)_0^{\rm mid-IR} = 18.96 \pm 0.01 \mbox{ (statistical)} \pm 0.03$
(systematic) mag (Scowcroft et al. 2015).

Although this is our final, recommended value based on the full body
of SMC distance estimates published during the past 2--3 decades, we
caution that an acute awareness of systematic effects possibly
exceeding 0.15--0.20 mag is of the utmost importance when using such a
generic mean value for practical purposes. Indeed, despite decades of
progress, we are still dealing with distance diagnostics that show
standard deviations of order 0.10 mag or more in the resulting
distance moduli. This means that more detailed analyses of both the
SMC's geometry and possible sources of systematic errors are still
urgently required (cf. Rubele et al. 2015; Ripepi et al., in prep.).

\begin{acknowledgements}
We thank Jan Lub for referring us to the early publication by
Hertzsprung, and James Wicker for providing us with his insights into
the intricacies of statistical analysis. RdG thanks his colleagues at
the Universidad de Concepci\'on (Chile), in particular Doug Geisler
and Michael Fellhauer, for their hospitality during the revision of
this paper. He is also grateful for research support from the National
Natural Science Foundation of China through grant 11373010. This work
was also partially supported by PRIN-MIUR (2010LY5N2T), `Chemical and
dynamical evolution of the Milky Way and Local Group galaxies' (PI
F. Matteucci). This research has made extensive use of NASA's
Astrophysics Data System Abstract Service. We thank the referee for
offering many insightful comments that made the conclusions reached in
this paper more robust.
\end{acknowledgements}

\end{document}